\documentclass[prd, superscriptaddress, tightenlines, longbibliography, nofootinbib, eqsecnum, amsfonts, amsmath, floatfix, notitlepage, twopage, twocolumn]{revtex4-1}
\pdfoutput=1

\usepackage[utf8]{inputenc}
\usepackage{mathrsfs}
\usepackage{euscript}
\usepackage{epsfig}
\usepackage{graphics}
\usepackage{graphicx}
\usepackage{amsmath}
\usepackage{amssymb}
\usepackage{bm}
\usepackage[usenames,dvipsnames,svgnames,table]{xcolor}
\usepackage{xspace}
\usepackage{wasysym}
\usepackage{times}
\usepackage{appendix}
\usepackage{lipsum}
\usepackage[nolist,nohyperlinks]{acronym}
\usepackage{float}
\usepackage{simplewick}
\usepackage{natbib, ifthen}
\usepackage{hyperref}

\defcitealias{Pratten:2016dsm}{PL16}

\newcommand{\orcidacronym}[2]{\href{http://orcid.org/#2}{#1}}
\newcommand{\ALorcid}{\orcidacronym{AL}{0000-0001-5927-6667}}

\newcommand{\la}{\langle}
\newcommand{\ra}{\rangle}

\newcommand{\mksym}[1]{\ifmmode {\rm #1}\else #1\fi}

\newcommand{\mnu}{\sum m_\nu}

\providecommand{\lea}{\la}
\providecommand{\gea}{\ga}

\providecommand{\alt}{\lea}
\providecommand{\agt}{\gea}
\providecommand{\text}[1]{\rm{#1}}

\newcommand{\Mpc}{\text{Mpc}}

\newcommand{\grad}{\nabla}

\newcommand{\Hunit}{\text{km}\,\text{s}^{-1}\,\Mpc^{-1}}

\providecommand{\muK}{\mu{\rm K}}
\providecommand{\arcmin}{{\rm arcmin}}

\newcommand{\eV}{\,\text{eV}}

\providecommand{\CAMB}{{\tt camb}}

\providecommand{\HALOFIT}{{\tt halofit}}

\newcommand{\begm}{\begin{pmatrix}}
\newcommand{\enm}{\end{pmatrix}}
\newcommand{\threej}[6]{{\begm #1 & #2 & #3 \\ #4 & #5 & #6 \enm}}

\newcommand\ba{\begin{eqnarray}}
\newcommand\ea{\end{eqnarray}}
\newcommand\bea{\begin{eqnarray}}
\newcommand\eea{\end{eqnarray}}

\newcommand\be{\begin{equation}}
\newcommand\ee{\end{equation}}

\newcommand{\valpha}{{\boldsymbol{\alpha}}}
\newcommand{\vgrad}{{\boldsymbol{\nabla}}}

\newcommand{\vell}{{\boldsymbol{\ell}}}

\newcommand{\ud}{{\rm d}}

\newcommand{\boldvec}[1]{{\mbox{\boldmath{$#1$}}}}

\newcommand{\vL}{\boldvec{L}}

\newcommand{\vd}{\boldvec{d}}
\newcommand{\ve}{\boldvec{e}}

\newcommand{\vk}{\boldvec{k}}
\newcommand{\vl}{\boldvec{l}}

\newcommand{\vn}{\boldvec{n}}

\newcommand{\vv}{\boldvec{v}}

\newcommand{\vx}{\boldvec{x}}

\newcommand{\clo}{\mathcal{O}}

\newcommand{\cls}{\mathcal{S}}

\newcommand{\vnhat}{\hat{\vn}}

\newcommand{\vkhat}{\hat{\vk}}

\renewcommand{\boldvec}[1]{{\boldsymbol{#1}}}
\newcommand{\isdraft}[1]{}

\newcommand{\AL}[1]{{\isdraft{\color{blue} AL: #1}}}
\renewcommand{\AC}[1]{{\isdraft{\color{green} AC: #1}}}

\newcommand{\trig}{\text{trig}}
\newcommand{\trigbar}{\overline{\text{trig}}}

\newcommand{\veps}{\boldvec{\epsilon}}

\newcommand{\parpropgeo}[2]{\Gamma_{#1}^{#2}}
\newcommand{\parproptrue}{\Gamma(\ve_\chi)_{\chi_\ast}^{0}}

\begin{document}


\newcommand{\calA}{{\mathcal{A}}}
\newcommand{\calC}{{\mathcal{C}}}
\newcommand{\calD}{{\mathcal{D}}}
\newcommand{\calR}{{\mathcal{R}}}
\newcommand{\calO}{{\mathcal{O}}}
\newcommand{\calS}{{\mathcal{S}}}
\renewcommand{\Re}{\operatorname{Re}}
\renewcommand{\Im}{\operatorname{Im}}
\def\n{\noindent}

\newcommand{\Sussex}{Department of Physics \& Astronomy, University of Sussex, Brighton BN1 9QH, UK}


\title{Emission-angle and polarization-rotation effects in the lensed CMB}
\author{Antony Lewis}
\affiliation{\Sussex}
\homepage{http://cosmologist.info}
\author{Alex Hall}
\address{Institute for Astronomy, University of Edinburgh, Royal Observatory, Blackford Hill, Edinburgh, EH9 3HJ, UK}
\author{Anthony Challinor}
\address{Institute of Astronomy and Kavli Institute for Cosmology, Madingley Road, Cambridge, CB3 0HA, UK}
\address{DAMTP, Centre for Mathematical Sciences, University of Cambridge, Wilberforce Road, Cambridge CB3 OWA, UK}

\begin{abstract}
Lensing of the CMB is an important effect, and is usually modelled by remapping the unlensed CMB fields by a lensing deflection. However the lensing deflections also change the photon path so that the emission angle is no longer orthogonal to the background last-scattering surface. We give the first calculation of the emission-angle corrections to the standard lensing approximation from dipole (Doppler) sources for temperature and quadrupole sources for temperature and polarization. We show that while the corrections are negligible for the temperature and E-mode polarization, additional large-scale B-modes are produced with a white spectrum that dominates those from post-Born field rotation (curl lensing). On large scales about one percent of the total lensing-induced B-mode amplitude is expected to be due to this effect. However, the photon emission angle does remain orthogonal to the perturbed last-scattering surface due to time delay, and half of the large-scale emission-angle B modes cancel with B modes from time delay to give a total contribution of about half a percent. While not important for planned observations, the signal could ultimately limit the ability of delensing to reveal low amplitudes of primordial gravitational waves.
We also derive the rotation of polarization due to multiple deflections between emission and observation.
The rotation angle is of quadratic order in the deflection angle, and hence negligibly small: polarization typically rotates by less than an arcsecond, orders of magnitude less than a small-scale image rotates due to post-Born field rotation (which is quadratic in the shear). The field-rotation B modes dominate the other effects on small scales.
\end{abstract}

\pacs{
}

\maketitle

\begin{acronym}
\acrodef{WL}[WL]{Weak Lensing}
\end{acronym}

\newcommand{\WL}{\ac{WL}\xspace}

\section{Introduction}

Lensing is the leading non-linear effect in the CMB because the distance to last scattering is much larger than the matter-radiation equality scale (so the large number of lenses boosts the amplitude), and because the small-scale perturbations and sharply-defined acoustic scale make the CMB quite sensitive to relatively small shifts in angular scale. For the B-mode polarization, lensing is also expected to be the dominant signal from scalar perturbations, and on large scales may also be an important confusing signal for any signal from primordial gravitational waves~\cite{Knox:2002pe,Kesden:2002ku,Hirata:2003ka}; see Ref.~\cite{Lewis:2006fu} for a review of CMB lensing.
The lensing B modes can be largely removed by delensing~\cite{Hirata:2003ka,Smith:2010gu,Carron:2017mqf}, though the extent to which this is possible in practice depends on the noise (and foregrounds), and the extent to which the simplest CMB lensing gradient-remapping approximation holds, as well as the size of other non-linear effects.

CMB lensing is usually modelled as a simple remapping, where the lensed CMB $\tilde X(\vnhat)$ observed in direction $\vnhat$ is taken to be what we would have observed in an unlensed CMB in direction $\vnhat'$, where the deflection angle $\valpha$ relates the two directions. This is, however, not strictly correct: because lensing changes the direction of photon propagation, a photon we observe originating from direction $\vnhat'$ will typically not have left the last-scattering surface on the same trajectory as it would have done in an unlensed universe. The photon emission angle is typically deflected away from the normal to the background last-scattering surface by an angle of $\clo(\alpha)$. This means that the treatment of sources with intrinsic angular dependence, i.e., the Doppler (dipole) source for the temperature anisotropies and the quadrupole source for the temperature and polarization, is in error at $\clo(\alpha)$ in the standard remapping approach. Since the angular dependence of the source is at most quadrupolar, and the deflection is $\clo(10^{-3})$, the effect is expected to be small: the change in emission angle does not change the observed scale of the perturbations, so is expected to be subdominant to the lensing convergence and shear, which enter with angular derivatives that enhance the effect on small scales~\cite{Hu:2001yq,Lewis:2006fu,Saito:2014bxa}. Although the emission-angle effect has, in principle, been included in some previous theoretical analyses~\cite{Fidler:2014zwa,Saito:2014bxa} there has not been an explicit calculation and quantification, so it remains important to quantify whether the effect is negligible or not.
We give the first explicit calculation and show that, although small, the large-scale B-mode signal could ultimately be important once the main lensing signal has been substantially removed by delensing.

In the lensed universe the background last-scattering surface is also perturbed by (Shapiro) time delay. Photons are emitted perpendicular to the perturbed surface, as required by Fermat's principle, but the source tensors have to be evaluated at the perturbed positions. The angular gradient of the time delay is the same as the emission angle to the background normal (as required for the emission direction to remain orthogonal to the perturbed surface), so both effects need to be considered together as they are parametrically of the same order. Reference~\cite{Hu:2001yq} has calculated the time-delay terms, so in this paper we focus mainly on the emission-angle corrections. However, to calculate the full correction to the lens-remapping approximation we include both effects, and show that for the polarization there are substantial cancellations between them. This reflects the fact that a locally-constant gradient in the last-scattering distance produces almost no effect on the power spectrum of small-scale perturbations because the local emission geometry looks like a rotated version of the background geometry (and hence has the same power by statistical isotropy).

For the polarization, there is also the possibility that the emitted polarization at last scattering gets rotated before it is observed.
Lensing by vector and tensor modes can rotate polarization at linear order~\cite{Dai:2013nda}, but here we focus on scalar perturbations.
In the Born approximation, interactions between multiple lensing events along the line of sight are neglected and so events are treated independently. For a single lensing deflection along the background line of sight, the entire photon path, the emission point and observed point all lie on a single plane.
By symmetry, for lensing by scalar perturbations, which have no handedness, there can be no polarization rotation in this case. However, at next order where the influence of lenses at different distances along the line are included on the action of any given lens, coplanarity can be lost and polarization rotation generated. The rotation arises because as the photon trajectory is bent, the polarization is parallel transported along the unit sphere defined by the photon propagation directions. As is well known, parallel transport around a closed curve on the sphere will generate a net rotation given by the spherical area of the enclosed region. Approximating the lensing as propagating the polarization directly from the emitted to the observed direction is in error by the area enclosed by the true path and the direct (great-circle) route. The rotation angle is therefore expected to be $\clo(\alpha^2)\sim \clo(10^{-6})$ and hence negligible. However, there has recently been some confusion over this issue, so we give an explicit calculation and demonstrate that it is indeed negligible as expected (contrary to the conclusion of Refs.~\cite{Marozzi:2016und,Marozzi:2016qxl}\footnote{References~\cite{Fabbian:2017wfp,Takahashi:2017hjr} only appear to agree with Refs.~\cite{Marozzi:2016und,Marozzi:2016qxl} because they assumed their result.}).

The polarization-rotation effect should not be confused with field rotation that appears in post-Born lensing and can also generate B-mode polarization~\cite{Cooray:2002mj,Hirata:2003ka,Pratten:2016dsm}.
The field rotation describes how a ray bundle is twisted as it propagates, generating a rotation of a small image. To see that the field and polarization rotation
are distinct effects it is sufficient to consider a simple example. Imagine two shearing lenses (at different redshift) centred on a background line of sight, with the axis of their ellipticities  misaligned. The combination of the misaligned shears will generate a second-order field rotation of a ray bundle centred on the background line of sight. However, the central photon in the ray bundle remains exactly unperturbed, and hence its polarization does not rotate at all. For arbitrarily small lenses the field rotation can be come arbitrarily large, but any polarization rotation is negligible as all the rays in the bundle are very close to the unperturbed background ray. The field and polarization rotation are both second order in perturbations, but the field rotation is quadratic in the shear rather than quadratic in the deflection, and hence is much larger on small scales.

This paper is organised as follows.
We start in Sec.~\ref{sec:geometry} by describing the Born-approximation lensing geometry and show how to calculate the relevant emission angle.
Then in Sec.~\ref{sec:TT} we calculate the effect on the temperature power spectrum from the change in the Doppler (dipole) source. (The effect from the smaller quadrupole source is calculated in Appendix~\ref{app:temppol}.)
Section~\ref{sec:polarization} describes the transport of the polarization, and how at the leading post-Born level polarization rotation can be generated. We then use these results in Sec.~\ref{sec:polrotation} to show that the polarization rotation power spectrum and the effect on the CMB are both negligible. In Sec.~\ref{sec:polemission} we focus on the more interesting emission-angle effects, and calculate the corresponding CMB power spectra. For simplicity we initially neglect time-delay terms (which have been calculated separately in Ref.~\cite{Hu:2001yq}), however they are highly anti-correlated to the emission-angle effect on the polarization. We consider the combined effect of emission angle and time delay in Sec.~\ref{sec:delay} in order to calculate the full correction to the standard lensing B-mode power spectrum.

Throughout we assume a flat $\Lambda$CDM cosmological model with purely adiabatic scalar perturbations evolving according to general relativity.
For numerical results we use power spectra from \CAMB~\cite{Lewis:1999bs}, with baryon density $\Omega_b h^2=0.02214$, dark matter density $\Omega_c h^2=0.127$, scalar perturbation power at $k=0.05\,\Mpc^{-1}$ of $A_s=2.118\times 10^{-9}$ with constant spectral index $n_s=0.965$, Hubble parameter $H_0 =66.88\,\Hunit$,
reionization optical depth $\tau=0.0581$,
 and one minimal massive neutrino with $\mnu=0.06\eV$~\cite{Aghanim:2016yuo}. Corrections to the matter power spectrum from non-linear growth are modelled by using \HALOFIT~\cite{Smith:2002dz,Takahashi:2012em}. Since the effects that we are calculating are small, it is not necessary to calculate them to high precision, so we use the flat-sky and Limber approximations for simplicity where appropriate.
The full-sky generalization is straightforward, and briefly presented in Appendix~\ref{FullSky} for our main B-mode result.
We work in the conformal Newtonian gauge using natural units with $c=1$, neglecting correlations between the lenses and the CMB. We only consider second- and higher-order effects related to lensing, taking the CMB anisotropy sources at recombination to be those in linear-theory in the Newtonian gauge evaluated at the background recombination conformal time $\eta_*$. Additional non-linear effects at recombination are generally expected to be smaller as the lensing deflections, emission angle and time delay gradients are enhanced by the large number of lenses along the line of sight to last-scattering.
Throughout, we denote 3D derivatives projected perpendicular to the line of sight by $\vgrad_\perp$, while covariant derivatives on the sphere are denoted by $\nabla_a$ or $\vgrad$.

\section{CMB lensing geometry}
\label{sec:geometry}

\begin{figure}[htp]
\includegraphics[width = 0.35\textwidth]{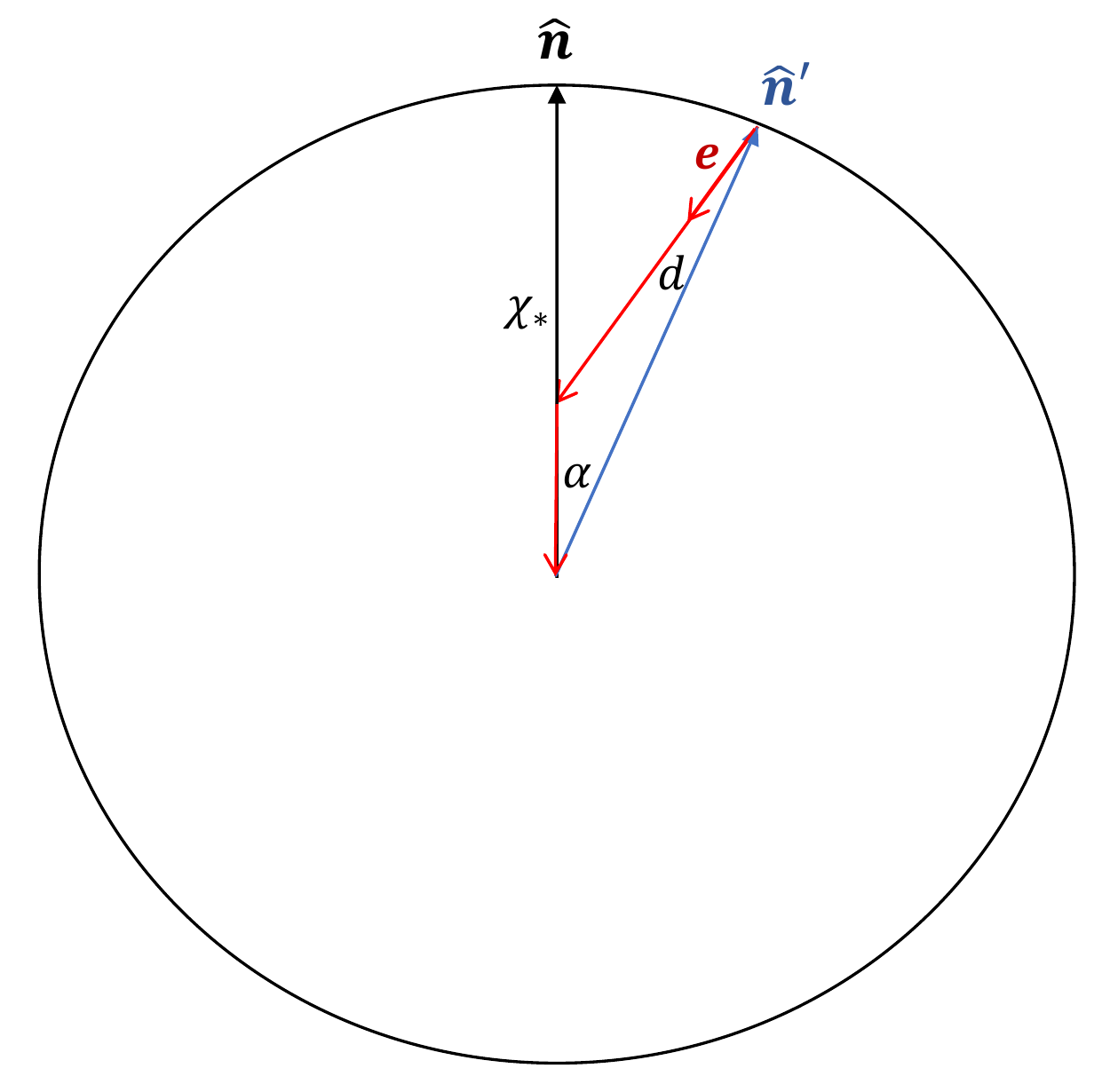}
\caption{Geometry of CMB weak lensing in the Born approximation. The photon observed in direction $\vnhat$ originates from direction $\vnhat'$ on the last-scattering surface, related by the deflection angle $\valpha$. The photon is emitted in the direction of the unit vector $\ve$, which differs from the normal $-\vnhat'$ by an angle $\vd$. The approximation usually used in CMB lensing is that the lensed and unlensed fields are related by $\tilde{X}(\vnhat) = {X}(\vnhat')$, where the unlensed field $X$ is evaluated in the background geometry so that it is assumed that the emission direction is $-\vnhat'$. In reality, the unlensed source differs from the background one because the components of the velocity, quadrupole and polarization tensors along $\ve$ differ at first order from those along $-\vnhat'$.
\label{firstordergeometry}
}
\end{figure}

\begin{figure}[htp]
\includegraphics[width = 0.45\textwidth]{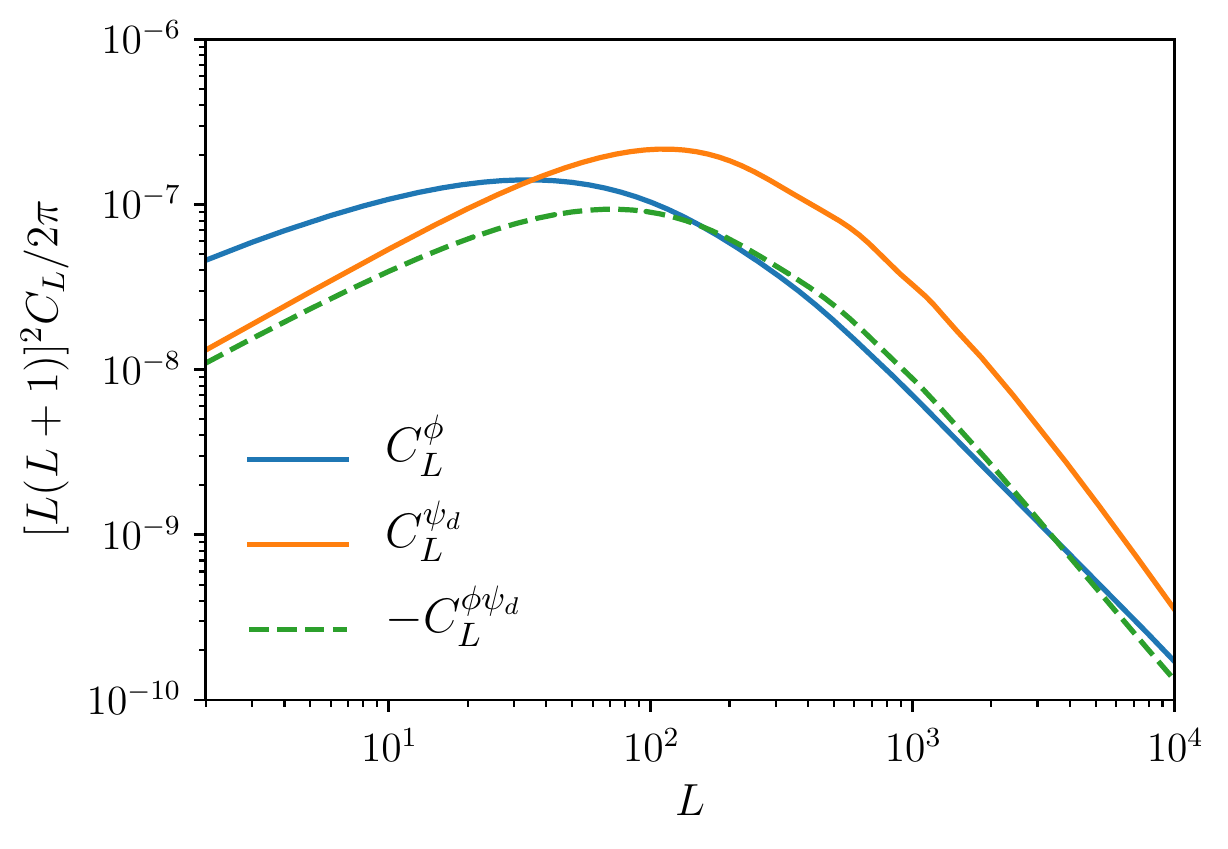}
\caption{Power spectrum of the standard lensing deflection $\valpha=\vgrad\phi$, compared to the power spectrum of $\vd=\vgrad\psi_d$, the deviation of the emission angle away from the normal to the background last-scattering surface. The sign is defined so that the cross-spectrum is negative: $\vd$ is defined to point in the direction of the transverse component of the emission direction, but $\valpha$ points from the background emission point to the actual emission point. Results are calculated using the Limber approximation and hence are not accurate at very low multipoles $L$.
\label{dpower}
}
\end{figure}

To start with we consider the lowest-order Born approximation, where all lensing events are evaluated along the background line of sight and interactions between these events are ignored, so they can be considered independently. The corresponding geometry for a single CMB lensing event is illustrated in Fig.~\ref{firstordergeometry}.
At emission, the photon direction $\ve$ makes an angle $d$ with the background direction $-\vnhat'$.
Analogously to the case of the deflection vector $\valpha$, we can define the tangent space vector $\vd(\vnhat)$ of length $d$ that points in the direction of the transverse component of $\ve$ (parallel transported back to $\vnhat$). In the flat-sky approximation $\vnhat'=\vnhat + \valpha$ and $\vd = \vnhat'+\ve = \valpha + \Delta\ve$, where $\Delta\ve$ is the difference between $\ve$ and $-\vnhat$. The lensing deflection angle between $\ve$ and the line of sight $\vnhat$ is calculated using the standard result $\ud \Delta\ve/\ud\chi = 2\vgrad_\perp \Psi$, and the result for $\valpha$ is the weighted form of this,
$\ud \valpha/\ud\chi = -2(1-\chi/\chi_*)\vgrad_\perp \Psi$. Here, $\Psi$ is the Weyl potential (equal to the Newtonian potential once radiation is negligible),
and $\chi_*$ is the comoving distance to recombination.


We define potentials so that for scalar perturbations $\valpha = \vgrad\phi$ and $\vd = \vgrad \psi_d$, where
   the relevant new power spectra in the Born and Limber approximations are then
\begin{eqnarray}
C_L^{\psi_d} &=& 4 \int_0^{\chi_*} \frac{\ud\chi}{{\chi}^2\chi_*^2}  P_\Psi(L_{\rm lim}/\chi,z(\chi)) \\
C_L^{\phi} &=& 4 \int_0^{\chi_*} \frac{\ud\chi}{{\chi}^4} (1-\chi/\chi_*)^2 P_\Psi(L_{\rm lim}/\chi,z(\chi)) \\
C_L^{\phi\psi_d} &=& -4 \int_0^{\chi_*} \frac{\ud\chi'}{{\chi}^3\chi_*}(1-\chi/\chi_*)P_\Psi(L_{\rm lim}/\chi,z(\chi)),\quad\,
\end{eqnarray}
where $P_\Psi$ is the power spectrum of the Weyl potential at the relevant redshift and $L_{\rm lim} \equiv L+1/2$.
The emission angle $d$ has an RMS amplitude of approximately $2.8\,\arcmin$, comparable to the deflection angle $\alpha$, but with more power on smaller scales, as shown in Fig.~\ref{dpower}.

\section{Emission-angle effect on the temperature power spectrum}
\label{sec:TT}

The standard lensing remapping approach is correct for scalar source terms for the temperature anisotropies. The most important non-scalar source is the Doppler effect from the electron bulk velocity at last scattering: $T_v(\vnhat) = e^{-\tau} \ve \cdot \vv_b$. Here, $\vv_b$ is the baryon velocity in the Newtonian gauge and $\tau$ is the optical depth to reionization. In the presence of lensing, the lensed Doppler contribution observed along the line of sight $\vnhat$ is
\begin{equation}
\tilde{T}_v(\vnhat) = e^{-\tau} \ve \cdot \vv_b(\chi_\ast \vnhat') ,
\end{equation}
where $\ve$ is the emission direction at the lensed point $\chi_\ast \vnhat'$ on the last-scattering surface. Writing $\ve = -\cos(d) \vnhat' + \sin(d) \hat{\vd}'$, where $\vd'(\vnhat')$ points along a great circle on the 2-sphere from $-\vnhat'$ to $\ve$ (i.e., $\vd(\vnhat)$ parallel transported to $\vnhat'$), we have
\begin{multline}
\ve \cdot \vv_b(\chi_\ast \vnhat') = -\cos(d) \vnhat' \cdot \vv_b(\chi_\ast \vnhat') \\
+ \sin(d) \hat{\vd}'(\vnhat') \cdot \vv_{b,\perp}(\vnhat';\chi_\ast \vnhat') ,
\end{multline}
where $\vv_{b,\perp}(\vnhat';\chi_\ast \vnhat')$ is the (transverse) projection of the baryon velocity perpendicular to $\vnhat'$ at the point $\chi_\ast \vnhat'$. This naturally decomposes the baryon velocity into a spin-0 field on the celestial sphere, $\vnhat\cdot \vv_b(\chi_\ast \vnhat)$, and a spin-1 field $\vv_{b,\perp}(\vnhat;\chi_\ast \vnhat)$. The evaluation of these fields at the lensed direction $\vnhat'$ can be handled with the usual covariant series expansion of the lensing-remapping approach~\cite{Challinor:2002cd}. For example, the spin-1 field $\vv_{b,\perp}(\vnhat';\chi_\ast \vnhat')$ after parallel transporting to $\vnhat$ along the connecting great circle is $\vv_{b,\perp}(\vnhat;\chi_\ast \vnhat) + \valpha(\vnhat)\cdot \vgrad
\vv_{b,\perp}(\vnhat;\chi_\ast \vnhat) + \cdots$. Note also that $\vd'(\vnhat')$ transports back to $\vd(\vnhat)$. The standard lensing-remapping approximation takes Doppler terms $\tilde T_{v}^{\rm std}(\vnhat) \approx -e^{-\tau}\vnhat'\cdot \vv_b(\chi_\ast \vnhat')$, so expanding and keeping terms to third order we find the correction term
\begin{multline}
\Delta \tilde{T} = e^{-\tau}\left( \frac{d^2}{2}   \vnhat\cdot \vv_b + \vd \cdot(\vv_{b,\perp} + \valpha\cdot\vgrad \vv_{b,\perp})+ \cdots\right),
\label{deltaTcorr}
\end{multline}
where everything is evaluated in direction $\vnhat$.
The first term is just the reduction in the usual dipole component because it is no longer observed along the background line of sight. The second term is the leading linear correction from the transverse velocity component, and the third term is the leading lensing correction to that.

\begin{figure}[tp]
\includegraphics[width = 0.45\textwidth]{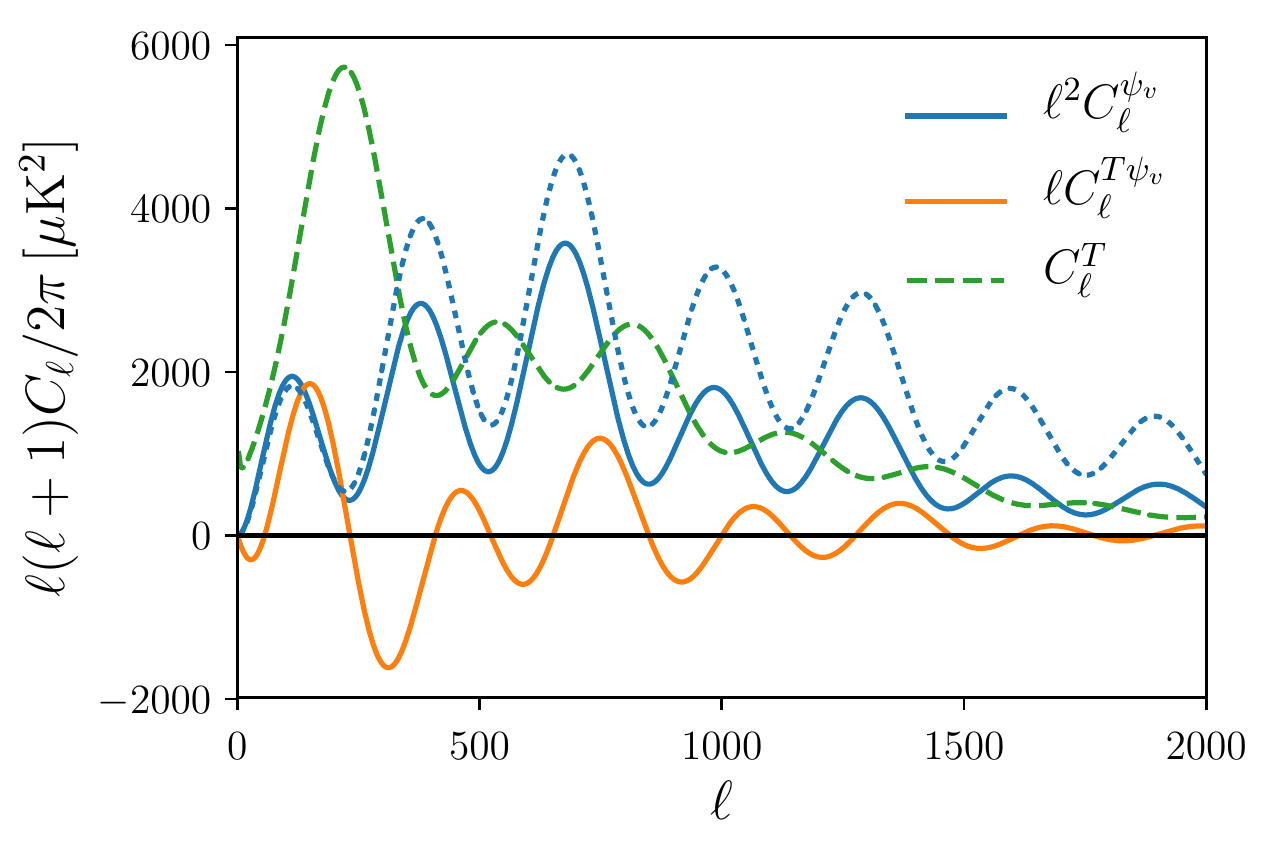}
\caption{Power spectrum of the transverse velocity potential from \CAMB\ with pre-reionization visibility weighting (solid blue)
and the delta-function visibility approximation of Eq.~\eqref{vel-CL-delta} (dotted blue).
The orange line shows the cross-correlation with the unlensed temperature (dashed green). Note that spectra are plotted with additional factors of $\ell$ so they all scale in roughly the same way, but the velocity potential power only enters with two fewer $\ell$ factors than shown here.
\label{velcl}
}
\end{figure}


The new terms depend on the
transverse component of the baryon velocity at recombination.
To give simple analytic results we use the delta-function visibility approximation, so the CMB photons are all sourced from a single sphere at comoving distance $\chi_*$.
Expanding in harmonics and defining a velocity potential $\psi_v$ we then have
\begin{align}
e^{-\tau} \vv_{b,\perp}(\vnhat;\chi_*\vnhat) &= \vgrad \psi_v(\vnhat) \\
&= - e^{-\tau}\vgrad \int \frac{\ud^3 \vk}{(2\pi)^3} \frac{v_b(\vk)}{k\chi_*} e^{i\vk\cdot \vnhat\chi_*},
\end{align}
where the transverse velocity potential has power spectrum
\begin{equation}
\label{vel-CL-delta}
C_\ell^{\psi_v} = 4\pi e^{-2\tau}\int \frac{\ud k}{k}  \frac{\mathcal{P}_v(k,\eta_*)}{(k\chi_*)^2} [j_\ell(k\chi_*)]^2.
\end{equation}
Here, $\mathcal{P}_v$ is the dimensionless power spectrum of $v_b(\vk)$.
For numerical work we calculate the power spectrum more accurately using \CAMB~\cite{Lewis:1999bs} to account for line-of-sight averaging due to the finite thickness of last scattering\footnote{The synchronous-gauge \CAMB\ source is $-(v^{\rm sync}_b+\sigma)g/(k\chi)$, where $g$ is the visibility before reionization and $\sigma$ the scalar part of the shear.}. The velocity potential is also correlated to the unlensed temperature anisotropies, and the corresponding numerical power spectra are shown in Fig.~\ref{velcl}.

\begin{figure}[tp]
\includegraphics[width = 0.45\textwidth]{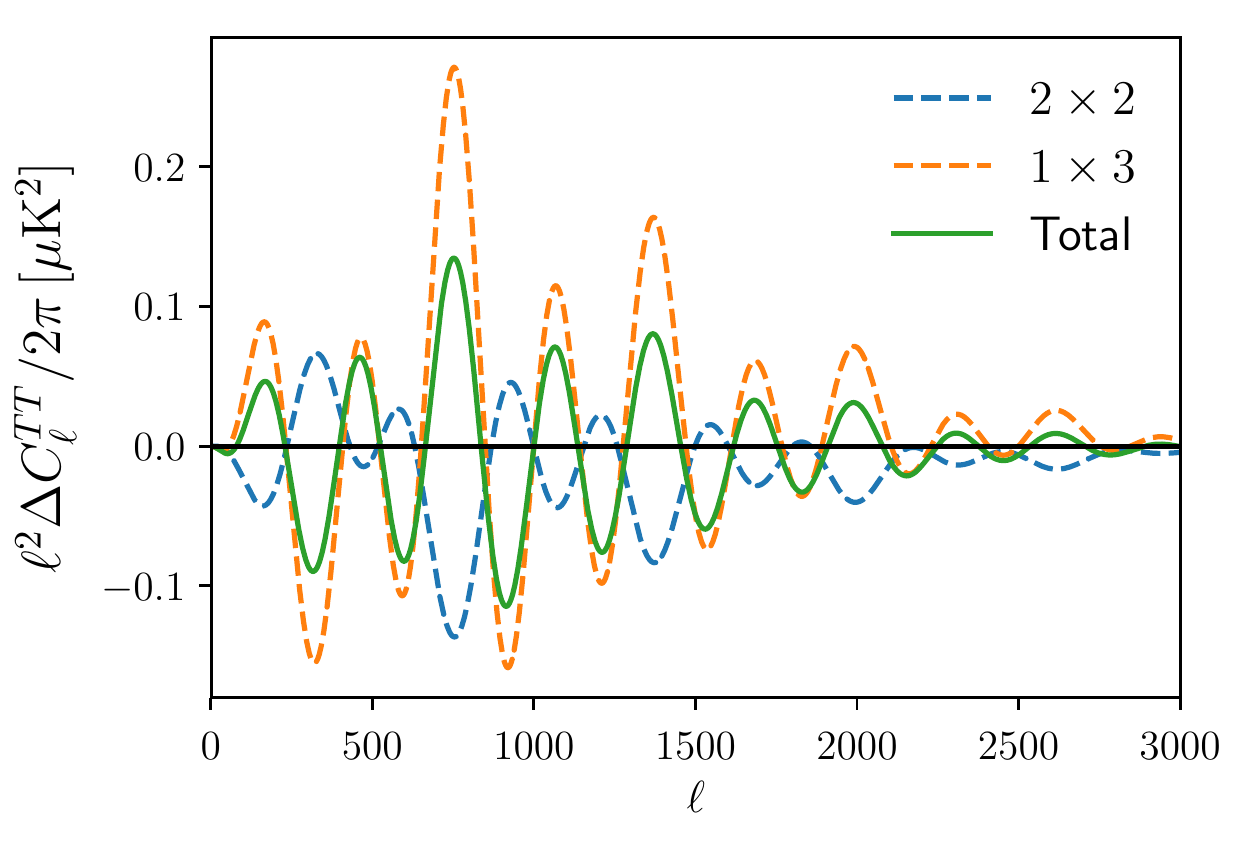}
\caption{Emission-angle effects on the CMB temperature power spectrum are dominated by terms involving the cross-correlation of the lensing and emission-angle Doppler terms. The convolution ($2\times 2$; blue) and $1\times 3$ (orange) terms in Eq.~\eqref{deltaTCL} are shown dashed and partly cancel, leaving the total shown in the solid green line.
\label{velTTauto}
}
\end{figure}

We write the total lensed temperature anisotropy as the standard lensing-remapping result plus the correction of Eq.~\eqref{deltaTcorr}. Series expanding to third order then gives
\begin{equation}
\tilde T = T + \valpha\cdot\vgrad T + \frac{1}{2}\alpha^a\alpha^b \grad_a \grad_b T + \Delta \tilde{T} + \cdots.
\end{equation}
Expanding into harmonics and calculating the power spectrum gives a correction to the standard perturbative result for the lensed temperature power spectrum given by
\begin{multline}
\Delta \tilde C_\ell^{TT} \approx \int \frac{\ud^2 \vL}{(2\pi)^2}  [\vL\cdot(\vell-\vL)]^2
\\ \hfill\times
\biggl( C_L^{\psi_d} C_{|\vell-\vL|}^{\psi_v}
+ 2C_L^{\psi_d\phi} C_{|\vell-\vL|}^{T\psi_v}
\biggr)\\
-\ell^2C_\ell^{T\psi_v}\int \frac{\ud L}{L} \frac{L^4 C_L^{\psi_d\phi}}{2\pi}
-C_\ell^{TT_v}\int \frac{\ud L}{L} \frac{L^4 C_L^{\psi_d}}{2\pi},
\label{deltaTCL}
\end{multline}
where $T_v(\vnhat) =-e^{-\tau}\vnhat\cdot\vv_b(\chi_\ast \vnhat)$ is the standard Doppler contribution to the unlensed temperature.
The first term in the convolution integral on the right-hand side is analogous to the usual ($2
\times 2$ order) lensing result, but is suppressed by $\clo(\ell^2)$ here as lensing brings in additional gradients [$\clo(\ell^2  C_{\ell}^{\psi_v})= \clo( C_\ell^{TT})$]: the contribution to the power is at the $10^{-3}\,\muK^2$ level, hence negligible. The second term in the convolution
integral comes from cross-correlation of the emission-angle effect with the leading term in the expansion of the standard lensing-remapping approximation, $\valpha\cdot\vgrad T$. This term is larger, as shown in Fig.~\ref{velTTauto}, because lensing involves additional transverse gradients. The remaining terms are $1\times 3$ order; their sum is similar in magnitude to the convolution term and partly cancel with it. The total is negligible compared to cosmic variance, so for the CMB temperature the standard lensing remapping approximation is adequate.

The temperature (and E-mode polarization) quadrupole at recombination also contributes to the observed temperature at a low level, so the perturbation to the emission angle will also affect the observed components of the quadrupole tensor. This is smaller than the Doppler effect, and hence negligible, but is discussed in Appendix~\ref{app:temppol} for completeness.

\section{Corrections to the polarization remapping approximation }
\label{sec:polarization}

The polarization observed along the line of sight $\vnhat$ was emitted at the lensed position $\chi_\ast \vnhat'$ in the propagation direction $\ve$.
The polarization source tensor is defined by $\zeta_{ab} \equiv \frac{3}{4}I_{ab} - \frac{9}{2}{\cal E}_{ab}$ and has contributions both from the temperature quadrupole $I_{ab}$ and the E-mode polarization quadrupole ${\cal E}_{ab}$ (see  Ref.~\cite{Challinor:2000as}). For convenience we define the polarization temperature source $\calS_{ab} \equiv -\zeta_{ab}/4$. The emitted linear polarization tensor is given by the transverse projection of the source tensor\footnote{Note that Ref.~\cite{Challinor:2000as} has a sign error that is corrected in Ref.~\cite{Tsagas:2007yx}.
In \CAMB\ the sign convention for the polarization is also opposite to that used here.}
\begin{equation}
P_{ab}^{\rm emit}(\ve) = [(\delta_{ac}-e_a e_c)(\delta_{bd}-e_{b} e_d)\calS_{cd}(\chi_*\vnhat', \eta_*)]^{\rm TT},
\end{equation}
where $[\ldots]^{\rm TT}$ denotes the symmetric, trace-free part projected orthogonal to $\ve$. As the photon direction $\ve(\chi)$ changes along the line of sight due to lensing, it describes a curve on the unit sphere connecting the emitted direction $\ve$ and the observed direction $-\vnhat$. As we review briefly below, the polarization tensor is parallel-transported along this curve. Denoting the operation of parallel transport along this path by $\parproptrue$, the observed, lensed polarization is given by\footnote{For analytic results we neglect reionization except via an overall $e^{-\tau}$ damping of the sub-horizon perturbations.}
\begin{equation}
\tilde{P}_{ab}(-\vnhat) = e^{-\tau}\parproptrue P_{ab}^{\rm emit}(\ve; \chi_\ast \vnhat') ,
\label{eq:obspol}
\end{equation}
where $P_{ab}^{\rm emit}(\ve; \chi_\ast \vnhat')$ is the emitted polarization along direction $\ve$ at the lensed position $\chi_\ast \vnhat'$.
The standard lens remapping result instead uses the polarization emitted at position $\chi_\ast \vnhat'$ along the radial direction $-\vnhat'$, and parallel transports this along the great circle connecting $\vnhat$ and $\vnhat'$. Denoting this parallel-transport operation by $\parpropgeo{-\vnhat'}{-\vnhat}$, the standard result is
\begin{equation}
\tilde{P}^{\rm std}_{ab}(-\vnhat) = e^{-\tau} \parpropgeo{-\vnhat'}{-\vnhat}
P_{ab}^{\rm emit}(-\vnhat'; \chi_\ast \vnhat') .
\label{eq:stdlens}
\end{equation}
This is in error because the true emission angle $\ve$ differs from $-\vnhat'$, and because the emitted polarization is parallel transported along a different path, with the latter effect giving rise to a rotation of the polarization. In this section we calculate these corrections to the standard lens remapping result.

Generally, the polarization of a photon is described by a (complex) spacelike 4-vector $\epsilon^\mu$ with $\epsilon_\mu^\ast \epsilon^\mu = -1$. In the Lorenz gauge, this is perpendicular to the photon 4-momentum $p^\mu$ and is parallel transported along the photon path in spacetime. A gas of photons can be described by a Hermitian-tensor-valued one-particle distribution function, $f_{\mu\nu}(x^\alpha,p^\beta)$, such that the number density of photons in phase space in a polarization state $\epsilon^\mu$ is $\epsilon^{\mu}{}^\ast f_{\mu\nu} \epsilon^\nu$. The tensor $f_{\mu\nu}$ is parallel transported along the photon path in phase space (Liouville's theorem).
For an observer with 4-velocity $u^\mu$, we can introduce the \emph{observed polarization vector} $\tilde{\epsilon}^\mu$ obtained by projecting $\epsilon^\mu$ with the screen-projection tensor $\mathcal{H}_{\mu\nu} = g_{\mu\nu} - u_\mu u_\nu + e_\mu e_\nu$:
\begin{equation}
\tilde{\epsilon}^\mu = \mathcal{H}^\mu{}_\nu \epsilon^\nu .
\end{equation}
Here, $g_{\mu\nu}$ is the spacetime metric and $e^\mu$ is a unit spacelike vector describing the propagation direction relative to the observer (so that $p^\mu \propto u^\mu + e^\mu$). The observed polarization has the virtue of being independent of the residual (electromagnetic) gauge freedom in the Lorenz gauge, which allows the addition to $\epsilon^\mu$ of a vector parallel to $p^\mu$. The dynamics of $\tilde{\epsilon}^\mu$ follows from the parallel transport of $\epsilon^\mu$:
\begin{equation}
\mathcal{H}^\mu{}_\nu (p^\alpha \nabla_\alpha \tilde{\epsilon}^\nu) = 0.
\label{eq:covtransport}
\end{equation}
We can similarly screen-project the distribution function to obtain $\tilde{f}_{\mu\nu} = \mathcal{H}_\mu{}^\alpha \mathcal{H}_\nu{}^\beta f_{\alpha\beta}$, which satisfies
\begin{equation}
\mathcal{H}_\mu{}^\rho \mathcal{H}_\nu{}^\tau \left(p^\alpha \nabla_\alpha \tilde{f}_{\rho\tau}\right) = 0 .
\end{equation}
It follows that the components of $\tilde{f}_{\mu\nu}$ are constant when expressed relative to a pair of screen-projected basis vectors that are transported according to Eq.~\eqref{eq:covtransport}. The symmetric, trace-free part of $\tilde{f}_{\mu\nu}$ is the linear polarization tensor $P_{ab}(\ve)$.

For scalar perturbations, we work in the Newtonian gauge
\begin{equation}
ds^2 = a^2(\eta) \left[(1+2\psi)d\eta^2 - (1-2\phi)\delta_{ij}dx^i dx^j\right] ,
\end{equation}
and introduce an orthonormal tetrad of vectors with $X_0 = a^{-1}(1-\psi)\partial_\eta$ and $X_i = a^{-1}(1+\phi)\partial_i$ for $i=1,2,3$. We adopt an observer with 4-velocity equal to $X_0$, i.e., at rest in the spatial coordinates. The photon direction has components $e^{\hat{\imath}}$ relative to this tetrad, which we denote by $\ve$. The observed polarization is similarly represented by its tetrad components $\tilde{\epsilon}^{\hat{\imath}}$, which we denote by the unit complex 3-vector $\tilde{\veps}$ with $\tilde{\veps}\cdot \ve = 0$. Expressed in terms of $\tilde{\veps}$, Eq.~\eqref{eq:covtransport} reduces to
\begin{equation}
\left(\delta_{ij} - e_{\hat{\imath}} e_{\hat{\jmath}}\right) \frac{d \tilde{\epsilon}^{\hat{\jmath}}}{d\eta} = 0.
\label{eq:3Dtransport}
\end{equation}
Along with the constraint $\tilde{\veps} \cdot \ve = 0$, this fully determines the evolution of the polarization direction. Equation~\eqref{eq:3Dtransport} has a simple physical interpretation: as the photon propagates, its direction $\ve(\eta)$ defines a curve on the unit sphere and the polarization direction $\tilde{\veps}$ is parallel transported along this curve. The polarization tensor $P_{ab}$ inherits the same parallel-transport law.

\begin{figure}[htp]
\includegraphics[width = 0.45\textwidth]{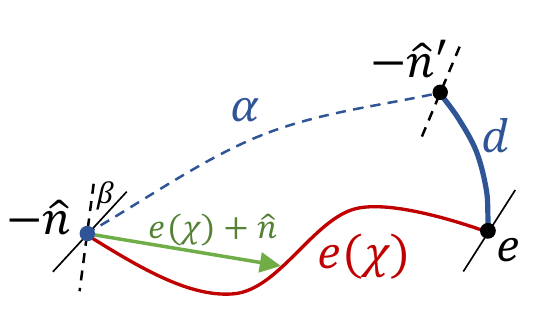}
\caption{%
Geometry of the transport of CMB polarization on a section of the sphere of photon propagation directions, with the polarization directions indicated by the black sticks. The polarization observed to be propagating along $-\vnhat$ is emitted along a direction $\ve$ (at the point $\chi_\ast \vnhat'$ on the last-scattering surface). The polarization is parallel transported along the path $\ve(\chi)$ on the sphere from $\ve$ to $-\vnhat$. The standard lens remapping result instead considers the polarization emitted along $-\vnhat'$ and parallel-transports this to the direction $-\vnhat$ along the great circle connecting these two directions. We relate the standard result to the true transport by first parallel transporting from $\ve$ to $-\vnhat'$ along the connecting great circle of length $d$. We further parallel transport from $-\vnhat'$ to $-\vnhat$ along the connecting great circle of length $\alpha$. The polarization after this latter two-step path is denoted by the dashed stick at $-\vnhat$. This has to be further rotated (in a right-handed sense) by an angle $\beta$ about $\vnhat$ to obtain the true polarization after propagation along $\ve(\chi)$ (denoted by the solid stick at $-\vnhat$). The magnitude of $\beta$ is given by the area enclosed by the path $\ve(\chi)$ and the great circles of length $d$ and $\alpha$; for the configuration shown here $\beta$ is positive. Generally, for small angles, $\beta$ is given by Eq.~\eqref{eq:finalbeta}.
\label{secondordergeometry}
}
\end{figure}

As the true curve $\ve(\eta)$ between the emission direction $\ve$ and the received direction $-\vnhat$ differs from the great circle that connects these directions, parallel transporting the polarization tensor along the true path will lead to a rotation compared to parallel transport along the great circle, with the angle given by the spherical area enclosed by the two paths. The geometry is illustrated in Fig.~\ref{secondordergeometry}. In detail, if transporting along $\ve(\eta)$ gives a rotation $\beta'$ (in a right-handed sense about $\vnhat$) compared to transport along the great circle, for small deflections
\begin{equation}
\beta' = \frac{1}{2}\vnhat \cdot \int_0^{\chi_\ast} \frac{d\ve}{d\chi} \times [\ve(\chi)+\vnhat] \, d\chi ,
\label{eq:betaprime}
\end{equation}
where we have switched to comoving distance $\chi$ for later convenience. This will be second order in the deflections, i.e., $\mathcal{O}(\alpha^2)$. Note that rotation is a post-Born effect: it relies on a non-zero $\ve(\chi)+\vnhat$ from lensing elsewhere along the line of sight interacting with the lensing event $d\ve / d\chi$ at $\chi$.

To relate to the standard remapping result, we also have to deal with the fact that the emission angle $\ve$ differs from the radial vector from the emission event, $-\vnhat'$. We do so as follows. First, at the lensed point $-\chi_\ast \vnhat'$ on the last-scattering surface, we parallel transport the emitted polarization tensor for the direction $\ve$ along the great circle from $\ve$ to $-\vnhat'$. This operation, which we denote by $\parpropgeo{\ve}{-\vnhat'}$, can be expressed as a covariant series expansion~\cite{Challinor:2002cd}
\begin{eqnarray}
\parpropgeo{\ve}{-\vnhat'} P_{ab}^{\rm emit}(\ve)
&=&  P_{ab}^{\rm emit}(-\vnhat') + \vd'\cdot\vgrad_{\ve}  P_{ab}^{\rm emit}(-\vnhat')
\nonumber\\ &&+
\frac{1}{2}d'^c d'^d \grad_{\ve,c} \grad_{\ve,d} P_{ab}^{\rm emit}(-\vnhat') + \cdots
\nonumber\\&=&
P_{ab}^{\rm emit}(-\vnhat')
+ 2 d'_{\langle a}\calS^{(1)}_{b\rangle}(\vnhat') - d'_{\la a} \calS_{b\ra c} d^{\prime c}
\nonumber\\ &&
+ d'_{\la a}d'_{b\ra}\hat n'^c \hat n'^d \calS_{cd} +\cdots .
\label{Pcovariantseries}
\end{eqnarray}
Here, $\vd'(\vnhat')$ is the tangent vector of length $d$ that points along a great circle towards $\ve$ (i.e., just $\vd(\vnhat)$ parallel transported to $\vnhat'$), angle brackets denote the symmetric, trace-free, projected part on the enclosed indices (i.e., the TT projection),
and the projected vector
$\calS_a^{(1)}(\vnhat) \equiv (\delta_a^b - \hat{n}_a \hat{n}^b)\calS_{bc}\hat{n}^c$. In Eq.~\eqref{Pcovariantseries}, the source tensor $\calS$ and $\calS_a^{(1)}(\vnhat')$ are evaluated at position $\chi_*\vnhat'$. Note that we can relate the projected $\cls_{ab}$ that appears in the $ d'_{\la a} \calS_{b\ra c} d^{\prime c}$ term to the (projected and trace-free) polarization tensor using the fact that $\calS_{ab}$ is 3D trace free, so Eq.~\eqref{Pcovariantseries} can also be written as
\begin{multline}
\parpropgeo{\ve}{-\vnhat'} P_{ab}^{\rm emit}(\ve) =
P_{ab}^{\rm emit}(-\vnhat')
+ 2 d'_{\langle a}\calS^{(1)}_{b\rangle}(\vnhat')
\\
- d'_{\la a} P^{\rm emit}_{b\ra c}(-\vnhat') d^{\prime c}
+ \frac{3}{2} d'_{\la a}d'_{b\ra}\hat n'^c \hat n'^d \calS_{cd} +\cdots .
\label{Pcovariantseriesalt}
\end{multline}
In this way, the contributions of the source tensor $\calS_{ab}$ have been decomposed into a spin-2 field (the TT projection $P^{\rm emit}_{ab}(-\vnhat; \chi_\ast \vnhat)$, where the second argument denotes the 3D position), a spin-1 field $\calS_a^{(1)}(\vnhat;\chi_\ast \vnhat)$, and a spin-0 field $\hat{n}^a \hat{n}^b \calS_{ab}(\chi_\ast\vnhat)$.

The first term on the right of Eq.~\eqref{Pcovariantseriesalt} is related to the standard lens-remapping result for the lensed polarization, Eq.~\eqref{eq:stdlens}, by parallel transport from $-\vnhat'$ to the observed direction $-\vnhat$ (and multiplication by $e^{-\tau}$).  Applying the $\parpropgeo{-\vnhat'}{-\vnhat}$ parallel transport operation to the other terms can be handled with a covariant series expansion around $\vnhat$, e.g.,
\begin{multline}
\parpropgeo{-\vnhat'}{-\vnhat} \calS^{(1)}_a(\vnhat';\chi_\ast \vnhat') =
\calS^{(1)}_a (\vnhat;\chi_\ast\vnhat) \\
+ \valpha \cdot \vgrad \calS^{(1)}_a (\vnhat;\chi_\ast\vnhat) + \cdots ,
\end{multline}
and $\vd'(\vnhat')$ transports back to $\vd(\vnhat)$. It follows that
\begin{multline}
e^{-\tau}\parpropgeo{-\vnhat'}{-\vnhat} \parpropgeo{\ve}{-\vnhat'} P_{ab}^{\rm emit}(\ve;\chi_\ast \vnhat') =
\tilde{P}_{ab}^{\rm std}(-\vnhat) \\
+ 2 e^{-\tau} d_{\langle a}\left[\calS^{(1)}_{b\rangle} + \valpha\cdot\vgrad  \calS^{(1)}_{b\rangle}\right]
- d_{\la a} P_{b\ra c}(-\vnhat) d^{c} \\
+ \frac{3}{2} e^{-\tau} d_{\la a}d_{b\ra}\hat n^c \hat n^d \calS_{cd} +\cdots ,
\label{eq:etonpton}
\end{multline}
correct to second order in lensing displacements. Here, all terms on the right are evaluated in direction $\vnhat$, and $P_{ab}(-\vnhat)$ is the unlensed polarization. The quantity on the left is almost the observed polarization, Eq.~\eqref{eq:obspol}, differing only because of the different paths involved in the parallel tranport and hence a rotation about $\vnhat$. To obtain the observed polarization, we rotate by an angle $\beta$, where $\beta$ is the area enclosed by the path $\ve(\chi)$ and the great circles connecting $\ve$ to $-\vnhat'$ and $-\vnhat'$ to $-\vn$; see Fig.~\ref{secondordergeometry}. This total area is the sum of $\beta'$ from Eq.~\eqref{eq:betaprime} and the area of the spherical triangle formed by $-\vnhat$, $\ve$, and $-\vnhat'$, so that for small angles to leading order we have
\begin{equation}
\beta = \frac{1}{2}\vnhat \cdot \int_0^{\chi_\ast} \frac{d\ve}{d\chi} \times [\ve(\chi)+\vnhat] \, d\chi + \frac{1}{2} \vnhat \cdot \left( \vd \times \valpha \right) .
\label{eq:finalbeta}
\end{equation}
Note that this is second-order in lensing deflections.

Expressing polarization in terms of complex combinations of the Stokes parameters,
${}_{\pm 2}P(\vnhat) = (Q\pm iU)(\vnhat) = e_\pm^{a} e_\pm^{b} P_{ab}(-\vnhat)$ (where, in the flat-sky approximation, $\ve_\pm= \ve_x\pm i\ve_y$ with $\ve_x$ and $\ve_y$ forming a right-handed set with the observed propagation direction $-\vnhat$), under a right-handed rotation of the polarization by $\beta$ about $\vnhat$ we have ${}_{\pm 2} P \rightarrow e^{\mp 2i\beta} {}_{\pm 2}P$. It follows from Eq.~\eqref{eq:etonpton} that the observed, lensed polarization is given to second order in deflections by
\begin{multline}
{}_{\pm 2}\tilde{P} =
e^{\mp 2i\beta}  {}_{\pm 2}\tilde P^{\rm std}+ e_\pm^{a} e_\pm^{b}\biggl[
2 e^{-\tau} d_{a}  (\calS^{(1)}_{b} + \valpha\cdot\vgrad \calS^{(1)}_{b} ) \\
- d_{a} P_{b c} d^{c}
+ \frac{3}{2} e^{-\tau} d_{a}d_{b}\hat n^c \hat n^d \calS_{cd}\biggr] +\cdots .
\label{fullpolrelation}
\end{multline}
The leading-order effect of rotation is just to rotate the usual lensed polarization by $\beta$, and the remaining terms are correct at third order in perturbations so there are no couplings between polarization rotation and emission angle terms. Since $\beta$ and $\psi_d$ are different parities, there are also no correlations between them.

\section{Polarization power from polarization rotation}
\label{sec:polrotation}

\begin{figure}[htp]
\includegraphics[width = 0.48\textwidth]{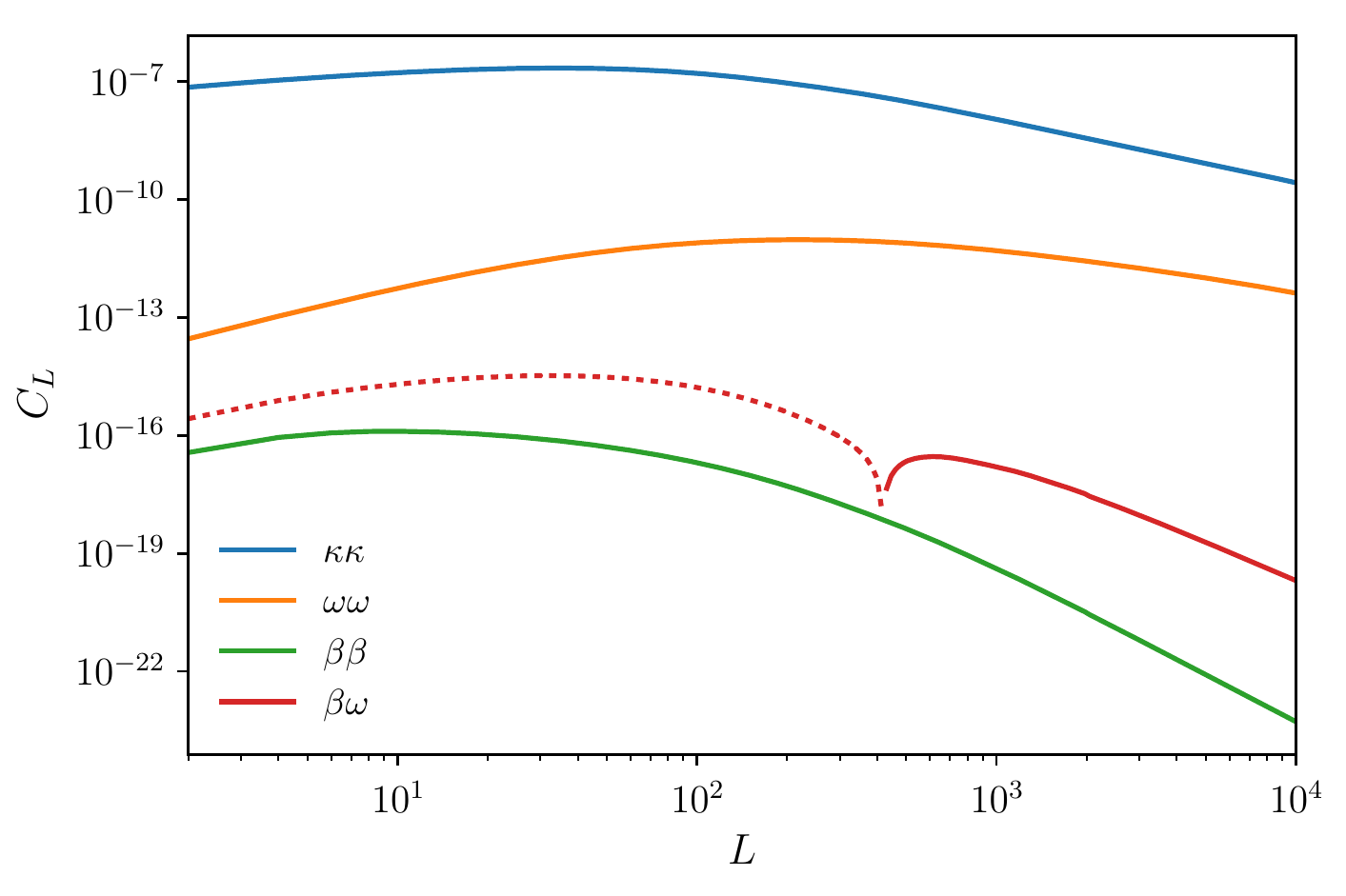}
\includegraphics[width = 0.48\textwidth]{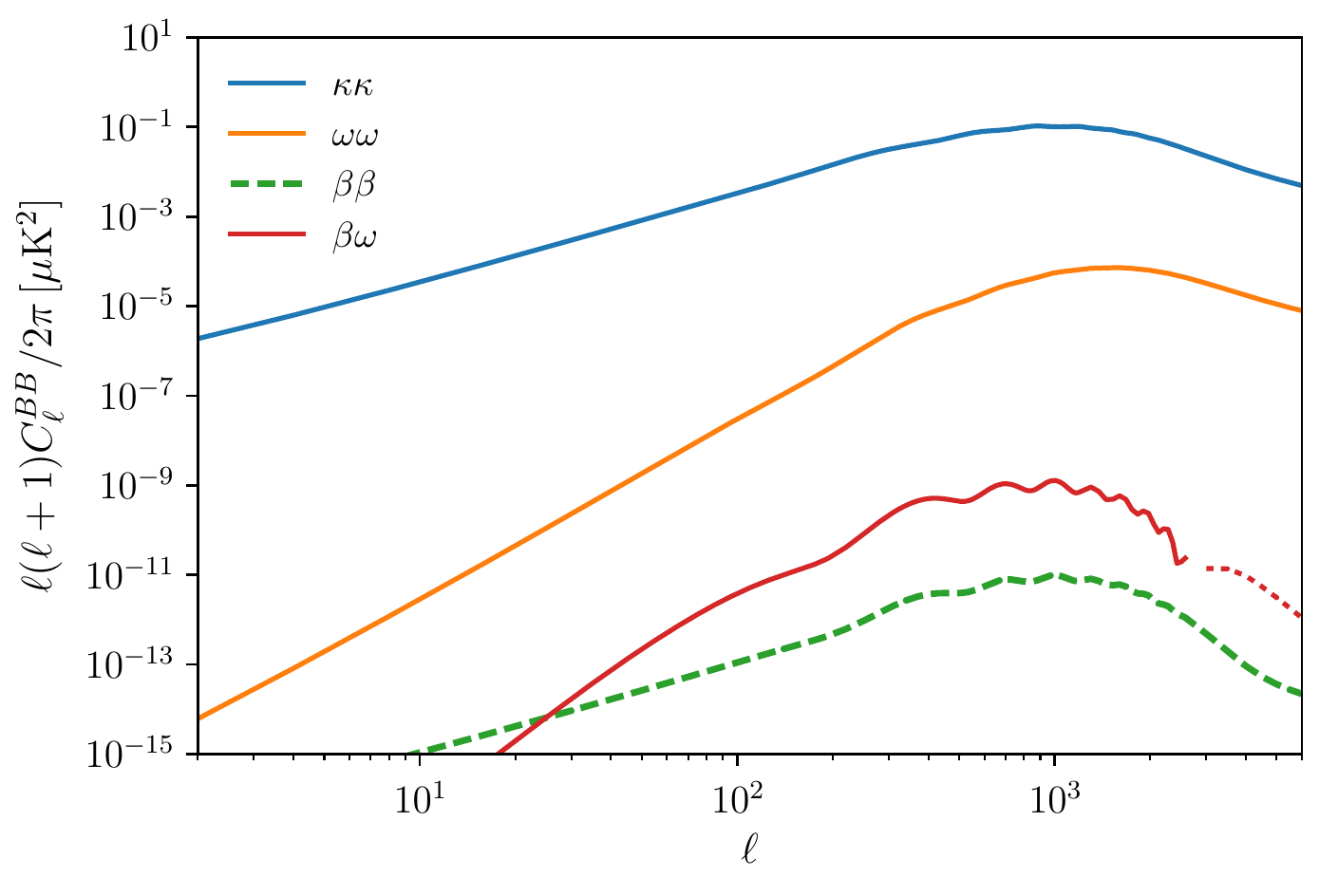}
\caption{\emph{Upper}: Power spectrum of the polarization rotation angle $\beta$ compared to the spectrum for the convergence $\kappa$ and field rotation $\omega$. The rotation and field rotation are both odd parity and are correlated (negative values shown as dotted lines).
\emph{Lower}: Corresponding $BB$ polarization power spectra. The polarization rotation contributions are negligible on all scales.
\label{powercomparequad}
}
\end{figure}

To lowest order we can evaluate the polarization rotation angle $\beta$ of the previous section
using the Born approximation results for $\ve(\chi)$ and $\valpha$. For the part $\beta'$, we have from Eq.~\eqref{eq:betaprime}
\begin{multline}
\beta' \approx 2 \epsilon_{ab} \int_0^{\chi_\ast} \ud\chi \, \nabla_\perp^a \Psi(\vnhat\chi, z(\chi))
\int_0^{\chi} \ud \chi' \\ \times \nabla_\perp^b\Psi(\vnhat\chi', z(\chi')) .
\end{multline}
For the contribution $\vnhat \cdot (\vd \times \valpha)/2$, we have
\begin{align}
\frac{1}{2} \vnhat \cdot (\vd \times \valpha) &= -2 \vnhat \cdot \left[ \int_0^{\chi_\ast} \ud \chi \,
\frac{\chi}{\chi_\ast} \vgrad_\perp \Psi(\vnhat\chi, z(\chi)) \right. \nonumber \\
&\quad \left. \times \int_0^{\chi_\ast}
\ud \chi' \, \left(1-\frac{\chi'}{\chi_\ast}\right) \vgrad_\perp \Psi(\vnhat\chi', z(\chi')) \right] \nonumber \\
&= -2 \epsilon_{ab} \int_0^{\chi_\ast} \ud\chi \, \nabla_\perp^a \Psi(\vnhat\chi, z(\chi))
\int_0^{\chi} \ud \chi'  \nonumber \\
&\qquad\qquad\qquad\nabla_\perp^b\Psi(\vnhat\chi', z(\chi')) \left(\frac{\chi-\chi'}{\chi_\ast}\right) ,
\end{align}
and combining with $\beta'$ gives
\begin{eqnarray}
\beta &\approx&  2\epsilon_{ab}\int_0^{\chi_*} \ud \chi
\nabla_\perp^a \Psi(\vnhat\chi, z(\chi))
\int_0^\chi \ud \chi'
\nonumber\\&&\qquad\times
\nabla_\perp^b \Psi(\vnhat\chi', z(\chi'))
\left(1-\frac{\chi}{\chi_*} + \frac{\chi'}{\chi_*}\right).
\end{eqnarray}
The corresponding lowest-Limber-approximation power spectrum is
\begin{align}
C_\ell^\beta &= 4  \int \frac{\ud^2\vL}{(2\pi)^2}
[\vL\times \vell]^2 \int_0^{\chi_*} \frac{\ud \chi}{\chi^4}
P_\Psi(L/\chi, z(\chi))
\nonumber\\&\quad\times
\int_0^\chi  \frac{\ud \chi'}{{\chi'}^4} P_\Psi\left( \frac{|\vell-\vL|}{\chi'},z(\chi')\right)
\left(1-\frac{\chi}{\chi_*} + \frac{\chi'}{\chi_*}\right)^2.
\end{align}
Numerically the RMS rotation is $\la \beta^2 \ra^{1/2}\approx 3\times 10^{-7}$, about a tenth of an arcsecond.
The rotation is quadratic in the $\clo(10^{-3})$ deflection angle and hence is very small. Indeed, it is much smaller than the post-Born field rotation $\omega$, which is quadratic in the shear and so much larger on small scales; see Fig.~\ref{powercomparequad}.

The odd-parity polarization rotation field is correlated to the field rotation. Using the standard series result for the field rotation~\cite{Hirata:2003ka,Pratten:2016dsm} we obtain
\begin{multline}
C_\ell^{\beta\omega} = 4  \int \frac{\ud^2\vL}{(2\pi)^2}
[\vL\times \vell]^2 \vL\cdot (\vell-\vL)\,
\\\times
\int_0^{\chi_*} \frac{\ud \chi}{\chi^4}\left(1-\frac{\chi}{\chi_*}\right) P_\Psi(L/\chi, z(\chi))
\int_0^\chi  \frac{\ud \chi'}{{\chi'}^4}\left(1-\frac{\chi'}{\chi}\right)
\\\times
\left(1-\frac{\chi}{\chi_*} + \frac{\chi'}{\chi_*}\right)P_\Psi\left( \frac{|\vell-\vL|}{\chi'},z(\chi')\right).
\label{omegabeta}
\end{multline}
Note that $\la \omega(\vnhat)\beta(\vnhat)\ra=0$ (first and second derivatives at a point are uncorrelated), so the cross-correlation spectrum changes sign to integrate to zero.

The higher-order nature and hence very small size of the rotation angle ensures that rotation effects are negligible,
but to be explicit we show some of the corresponding contributions to the CMB $BB$ power spectra in Fig.~\ref{powercomparequad}.
Here, we only considered terms involving the $\beta$ rotation power and $\omega$ field-rotation cross-correlation, neglecting cross-terms with the curl component of $\vd$ and using the Gaussian approximation results for the lensed power given in the Appendix~\ref{app:rotpower}.
The rotation-induced polarization is negligible on all scales, and has power spectrum that is parametrically smaller than the field-rotation induced signal since the latter scales like shear squared rather than deflection squared (hence scaling with a relative $\ell^2$).

Note that we have assumed lensing by purely scalar fluctuations. Non-linear evolution of scalar modes can also source secondary vector and tensor modes that could also rotate the polarization as well as produce field rotation~\cite{Dai:2013nda}. However, the sourced second-order non-scalar modes are also expected to have very low amplitude~\cite{Mollerach:2003nq,Baumann:2007zm,Sarkar:2008ii,Saga:2015apa,Adamek:2015mna}.

\section{Polarization power from emission-angle effects}
\label{sec:polemission}

\begin{figure}[tp]
\includegraphics[width = 0.45\textwidth]{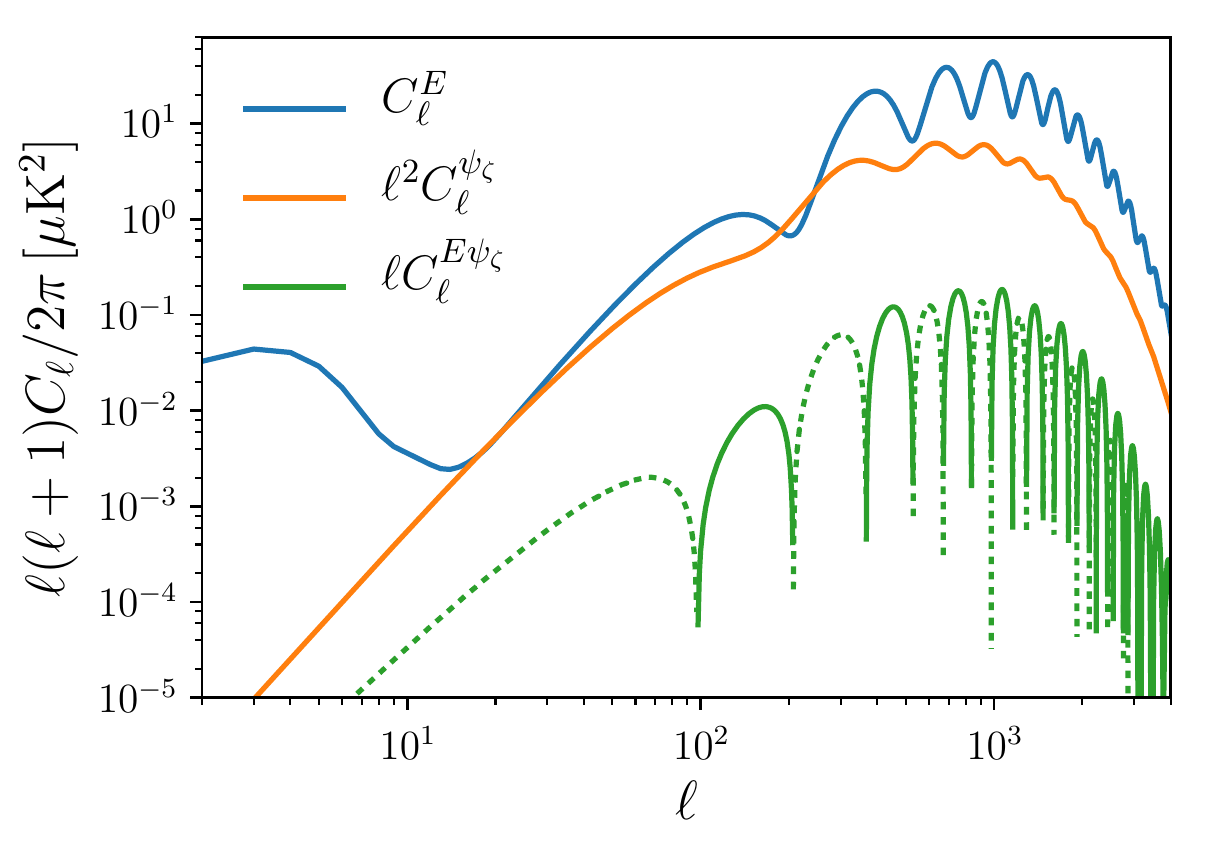}
\caption{Power spectrum of the longitudinal polarization potential from \CAMB\ (orange) and its cross-spectrum with the polarization potential $\psi_E$ (green, with dotted lines indicating negative values). Note that these spectra are shown with additional factors of $\ell$ to have similar scaling to $C^{EE}_{\ell}$ (blue, which here includes the additional large-angle power generated by scattering at reionization).
\label{polcl}
}
\end{figure}

Since the polarization rotation is negligible, we focus on the leading emission-angle terms of Eq.~\eqref{fullpolrelation} and revert to the Born approximation\footnote{
Post-Born corrections to $\vd$, which could enter the third-order $P_{ab}^{(3)}$ term discussed below,
would not contribute to the $1\times 3$ power spectrum term below because $\langle \vd \rangle=0$ by isotropy.
}. For scalars perturbations we can expand the symmetric, trace-free polarization source tensor in harmonics as
\begin{equation}
\calS_{ab} = -\int \frac{\ud^3 \vk}{(2\pi)^3} \left(\vkhat_a \vkhat_{ b} - \frac{1}{3}\delta_{ab}\right) \cls(\vk) e^{i\vk\cdot\vx}.
\end{equation}
The leading correction to the lensed polarization that is linear in $\vd(\vnhat)$ is then
\begin{eqnarray}
\Delta \tilde P^{(2)}_{ab} &=&
2e^{-\tau} d_{\la a} \calS^{(1)}_{b\ra}
\nonumber\\&=&
-\frac{1}{2}e^{-\tau}\int \frac{\ud^3 \vk}{(2\pi)^3} \frac{\zeta(\vk)}{k^2} i\vnhat\cdot\vk\,  d_{\la a} \grad_{\perp b\ra}  e^{i\vk\cdot\vx}
\nonumber\\&=&
-\frac{1}{2}e^{-\tau}d_{\la a}\grad_{b\ra} \int \frac{\ud^3 \vk}{(2\pi)^3} \frac{\zeta(\vk)}{k^2}\frac{\ud}{\ud \chi}
\left(\frac{  e^{i\vk\cdot\vx}}{\chi}\right)\quad
\label{deltaP}
\end{eqnarray}
evaluated at $\vx=\chi_*\vnhat$. Here, $\zeta(\vk) = -4 \cls(\vk)$.
We define the polarization potential
\begin{equation}
\psi_\zeta(\vnhat) \equiv
-\frac{1}{4}
e^{-\tau}\int \frac{\ud^3 \vk}{(2\pi)^3} \frac{\zeta(\vk)}{k^2}\frac{\ud}{\ud \chi_*} \left(\frac{e^{i\vk\cdot\vnhat\chi_*}}{\chi_*}\right),
\label{psizeta}
\end{equation}
so that $e^{-\tau}\calS^{(1)}_a = \nabla_a \psi_\zeta$ and $\Delta \tilde P_{ab}^{(2)}  = 2\grad_{\la a} \psi_d \grad_{b\ra} \psi_\zeta$. We further define
\begin{equation}
\psi_E(\vnhat) \equiv
-\frac{1}{4}
e^{-\tau}\int \frac{\ud^3 \vk}{(2\pi)^3} \frac{\zeta(\vk)}{k^2\chi_*^2} e^{i\vk\cdot\vnhat\chi_*},
\label{psiE}
\end{equation}
so that the unlensed polarization is $P_{ab} = \grad_{\la a} \grad_{b\ra}\psi_E$.
We define our sign convention for the flat-sky E and B harmonics so that the E-mode is $E(\vell)=\ell^2\psi_E(\vell)$ following Ref.~\cite{Lewis:2006fu}. When we write the unlensed E-mode field it should be understood as evaluated using only the sources at recombination (we discuss additional contributions from lensing of reionization sources in Sec.~\ref{sec:delay} below).

\begin{figure}[htp]
\includegraphics[width = 0.45\textwidth]{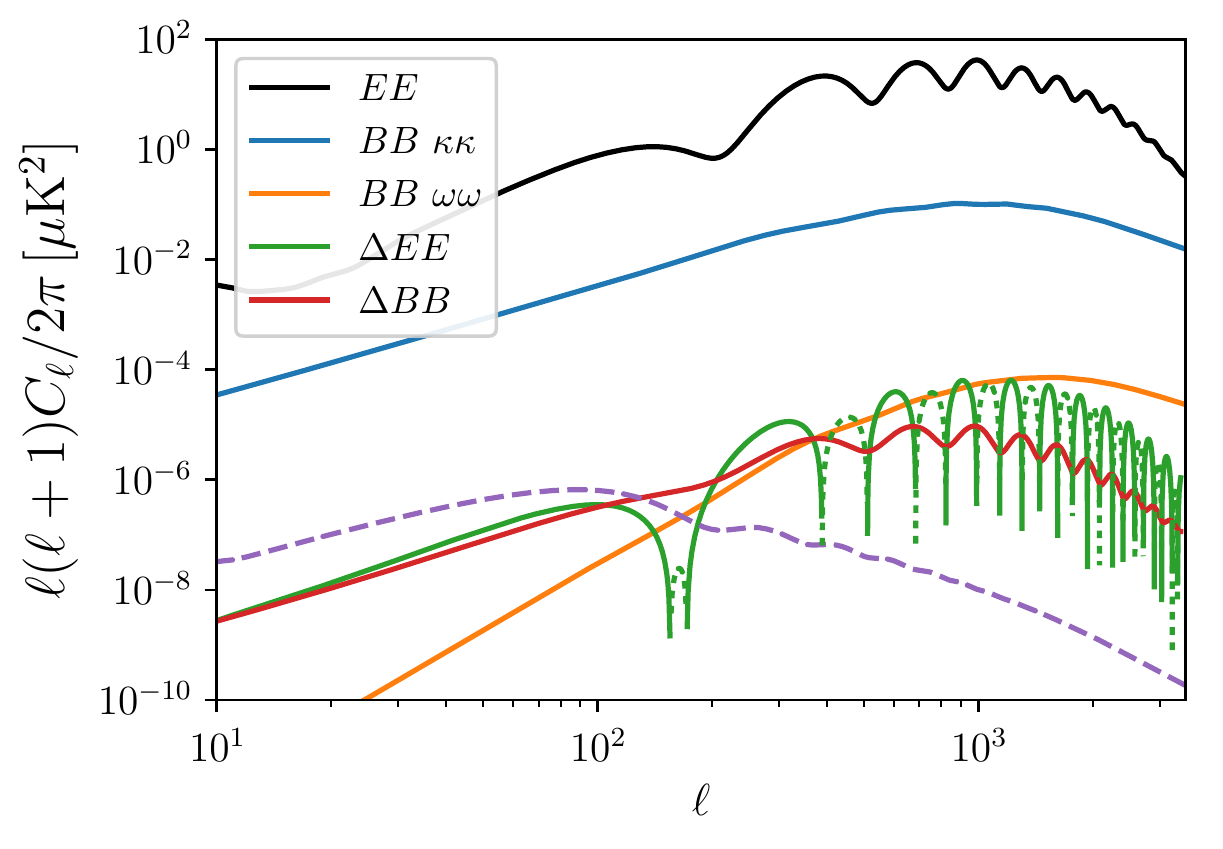}
\caption{
Emission-angle contributions to the CMB B-mode polarization power spectrum compared to standard gradient (blue) and post-Born curl (orange) lensing signals. The correction to the B-mode power spectrum from emission-angle effects (red) has a white spectrum on large scales, and,
on the scales most relevant for primordial gravitational waves, dominates that from field rotation
(the curl-induced B-mode polarization is a total derivative and has a blue spectrum on large scales).
For reference the dashed line shows the B-mode power expected from primordial gravitational waves with tensor-to-scalar ratio $r=10^{-5}$. The standard lensed E-mode spectrum is also shown (black) along with the corrections from emission-angle effects (green).
\label{poldeltacl}
}
\end{figure}

The corresponding relevant auto- and cross-spectra are
\begin{equation}
\label{vel-CL-delta-auto}
C_\ell^{\psi_\zeta} = \frac{\pi}{4} e^{-2\tau}\int \frac{\ud k}{k} \mathcal{P}_\zeta(k,\eta_*) \left[\frac{1}{k}\frac{\ud}{\ud \chi_*} \left(\frac{j_\ell(k\chi_*)}{k\chi_*}\right)\right]^2,
\end{equation}
\begin{multline}
\label{vel-CL-delta-cross}
C_\ell^{E\psi_\zeta} = \frac{\pi}{4}\ell^2 e^{-2\tau} \!\int \frac{\ud k}{k}  \mathcal{P}_\zeta(k,\eta_*)
\\
\times \frac{j_\ell(k\chi_*)}{{(k\chi_*)^2} }\frac{1}{k}\frac{\ud}{\ud \chi_*} \left[\frac{j_\ell(k\chi_*)}{k\chi_*}\right].
\end{multline}
Figure~\ref{polcl} shows corresponding numerical results from \CAMB\ using visibility-weighted sources.\footnote{The \CAMB\ source for $\psi_\zeta$ is $\frac{1}{4k^2\chi}\frac{\ud}{\ud \eta}(g \zeta)$ where $g$ is the visibility prior to reionization (with sign difference due to the sign convention of \CAMB).
}
Taking the spin-$\pm 2$ components of $\Delta \tilde{P}_{ab}^{(2)} $ and expanding in flat-sky spin harmonics
gives second-order terms including emission-angle and standard remapping effects
\begin{multline}
\Delta E(\vell) = \int \frac{\ud^2\vL}{(2\pi)^2} \biggl(2[\vell\cdot \vL - L^2 \cos(2\varphi_{L\ell})]
\psi_{\zeta}(\vL) \psi_d(\vell-\vL)
\\- \vL\cdot(\vell-\vL)\cos (2\varphi_{L\ell}) E(\vL) \phi(\vell-\vL)
\biggr),
\label{Edelta}
\end{multline}
\begin{multline}
\Delta B(\vell) = \int \frac{\ud^2\vL}{(2\pi)^2} \biggl(2[\vell\times \vL - L^2 \sin(2\varphi_{L\ell})]
\psi_{\zeta}(\vL) \psi_d(\vell-\vL)
\\- \vL\cdot(\vell-\vL)\sin (2\varphi_{L\ell}) E(\vL) \phi(\vell-\vL)
\biggr).
\label{Bdelta}
\end{multline}
Here, $\varphi_{L\ell}\equiv \varphi_L-\varphi_\ell$ is the angle between $\vL$ and $\vell$,
and we use $\vl \times \vL = \ell L \sin\varphi_{L\ell}$.
Evaluating the corresponding correction to the usual lens-remapping perturbative result for the power spectrum gives
\begin{multline}
\Delta \tilde C^{BB}_\ell = 4\int \frac{\ud^2\vL}{(2\pi)^2}\biggl( [\vell\times \vL - L^2 \sin(2\varphi_{L\ell})]^2 C^{\psi_{\zeta}}_L C^{\psi_d}_{|\vell-\vL|}
\\-[\vell\times \vL - L^2 \sin(2\varphi_{L\ell})]\vL\cdot(\vell-\vL)\sin (2\varphi_{L\ell})C^{E\psi_{\zeta}}_L C^{\phi\psi_d}_{|\vell-\vL|}
\biggr).
\label{CBemission}
\end{multline}
Numerical results are shown in Fig.~\ref{poldeltacl}.
On large scales the spectrum is white with
\begin{eqnarray}
\Delta \tilde C^{BB}_\ell &\approx& 4\int \frac{ L^5\ud L}{4\pi}\biggl( C^{\psi_{\zeta}}_L C^{\psi_d}_L
-C^{E\psi_{\zeta}}_L C^{\phi\psi_d}_L\biggr) \nonumber\\
&\approx& 2\times 10^{-10}\, \mu{\rm K}^2.
\label{BBemissionWhite}
\end{eqnarray}
The auto-correlation part (first term in brackets in the integrand) dominates the total, corresponding to an effective noise level of approximately $0.05\,\mu{\rm K}\,{\rm arcmin}$: about $1\,\%$ of the large-scale lensing-related B-mode amplitude should be due to the emission-angle effect.

To calculate the leading correction to the E-mode polarization power spectrum we need to include the third-order terms from Eq.~\eqref{fullpolrelation}:
\begin{multline}
\Delta \tilde P^{(3)}_{ab} \approx
2  e^{-\tau} d_{\la a}  \valpha\cdot \vgrad  \calS^{(1)}_{b\ra}
-d_{\la a} P_{b\ra c} d^c \\
+ \frac{3}{2} e^{-\tau} d_{\la a}d_{b\ra}\hat n^c \hat n^d \calS_{cd}.
\end{multline}
The first term $2\nabla_c \phi \grad_{\la a} \psi_d \nabla^c \nabla_{b\ra} \psi_\zeta$ comes from evaluating the leading emission-angle term at the lensed position $\chi_\ast \vnhat'$,
the second term accounts for reduction in the standard transverse polarization because it is no longer observed along the background line of sight,
and the third term is a new signal from sensitivity to the radiation quadrupole at last scattering aligned along the background line of sight.
The last term does not contribute to the $1\times 3$ power spectrum terms since
$\langle d_{\la a} d_{b\ra} \rangle = 0$ at a point,
but the other two terms do giving a total leading-order correction to the lensed power spectrum
\begin{multline}
\Delta \tilde C^{EE}_\ell = 4\int \frac{\ud^2\vL}{(2\pi)^2}
\biggl(
[\vell\cdot \vL - L^2 \cos(2\varphi_{L\ell})]^2 C^{\psi_{\zeta}}_L C^{\psi_d}_{|\vell-\vL|}
\\-[\vell\cdot \vL - L^2 \cos(2\varphi_{L\ell})]\vL\cdot(\vell-\vL)\cos (2\varphi_{L\ell})C^{E\psi_{\zeta}}_L C^{\phi\psi_d}_{|\vell-\vL|}
\biggr)
\\
+2\ell^2C_\ell^{E\psi_\zeta}\int \frac{\ud L}{L} \frac{L^4 C^{\phi\psi_d}_L}{2\pi}
-C_\ell^{EE}\int \frac{\ud L}{L} \frac{L^4 C^{\psi_d}_L}{2\pi}.
\end{multline}
 As shown in Fig.~\ref{poldeltacl}, this correction is many orders of magnitude smaller than the standard lensing result, and hence entirely negligible compared to cosmic variance.

\section{Time delay}
\label{sec:delay}

\begin{figure}[tp]
\includegraphics[width = 0.45\textwidth]{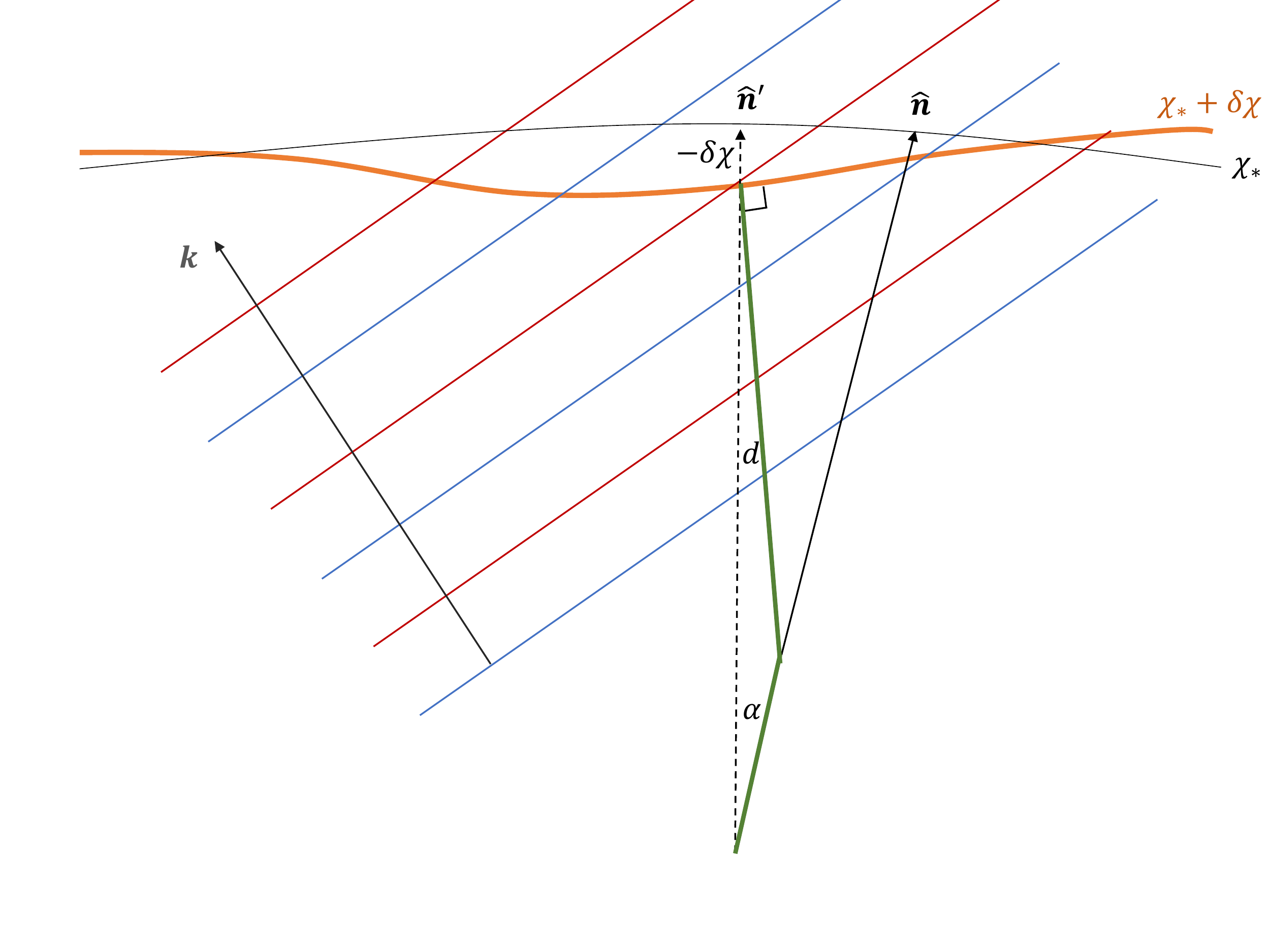}
\caption{The geometry of Born-approximation CMB lensing accounting for time delay, which distorts the background last-scattering surface (black arc of circle) by a radial distance $\delta\chi$ to give a perturbed surface (thick orange line).
Photons are emitted at angle $d$ to the normal $-\vnhat'$ to the background last-scattering surface, but are emitted perpendicular to the perturbed surface.
Red and blue lines indicate peaks and troughs of a plane-wave perturbation, with wavevector $\vk$, which describes the radiation perturbations.
If the last-scattering surface is at a nearly constant angle to the background line of sight compared to the wavelength of the radiation perturbations, the local photon emission geometry would be the same as in the background if $\vk$ were rotated by angle $d$ so that the emission trajectory, last-scattering surface and wavevector remain at the same angles
(and the phase is shifted to compensate the change in emission point). Since a rotated $\vk$ is statistically equivalent by isotropy, this means that any change in the small-scale power due to emission angle and time delay is suppressed (compared to either effect individually, where rotating the wavevector does not give the same geometry because the emission angle and last-scattering surface are then not orthogonal).
\AC{Could alternatively phrase this in terms of a shift in the position of the observer by $\chi_\ast \vd$ and looking along line-of-sight displaced by $\vd$? Clear then that you get strongly suppressed B-modes without invoking a statistical argument.}
\AL{This would be equivalent to saying the local emission geometry is the same as in the remapping approx, so no extra B generation?
But it doesn't explain more generally why the full $P_{ab}$ seems to get the 2nd order cancellation (is not sensitive to distance gradients except via the first term); though any more general explanation must invoke the scalar-mode assumption, which is what relates change under angle to change under displacement.
The integral-by-parts term cancellation argument could of course also be a bit misleading.
}
\AL{This is a statistical argument and not directly consistent with the map-level cancellation in the text. Perhaps we can say this:}
For B modes part of this cancellation happens at the map level at second order: the rotated emission geometry is locally equivalent to that in the standard remapping approximation and hence does not generate additional B modes from E modes.
\label{fig:timedelay}
}
\end{figure}

In the previous sections we have assumed that the source tensors are evaluated at the background last-scattering surface. However, as previously studied in Ref.~\cite{Hu:2001yq}, the sources are actually radially perturbed due to the time delay induced by potentials along the line of sight.
Including time delay, the CMB sources should be evaluated at $\vnhat'(\chi_*+\delta\chi)$, where $\delta\chi=2 \int_0^{\chi_*} \Psi \, \ud \chi $ is the time-delay correction to the background last-scattering radius. We neglect the geometric time delay as it is higher order, being second order in deflections.
The time delay is related to the emission angle (with potential $\psi_d$) because the geometric factors are such that $\delta\chi/\chi_* = \psi_d$. We can simply therefore reuse previous results for the emission angle power spectrum.
Note that the perturbation to the normal of the last-scattering surface is $\vgrad \delta\chi/\chi_* = \vgrad\psi_d = \vd$, so
the time delay distorts the last-scattering surface in just such a way that the lensed photon path remains normal to it.
Figure~\ref{fig:timedelay} illustrates the geometry, and also explains why we might expect to see a partial cancellation between the polarization signal produced by emission angle and time delay effects.

We have shown that the emission-angle terms are negligible for the temperature and E-mode polarization, and Ref.~\cite{Hu:2001yq} also computed the negligible time-delay terms. Both are dominated by the correlation with the lensing signal, so the cross-correlation of time-delay and emission-angle corrections will also be negligible, having one fewer derivatives of CMB fields. Here, we focus on the B-mode polarization where the small contributions could potentially become important if delensing can be applied efficiently. As we shall see, the time-delay B modes are highly anti-correlated with the emission-angle B modes, and they are parametrically of the same order, so it is important to consider both together to calculate the total accurately.
We have already shown that polarization rotation is negligible, so do not consider it further here.
e
\subsection{Total correction to the lensed B modes}

\begin{figure}
\includegraphics[width = 0.45\textwidth]{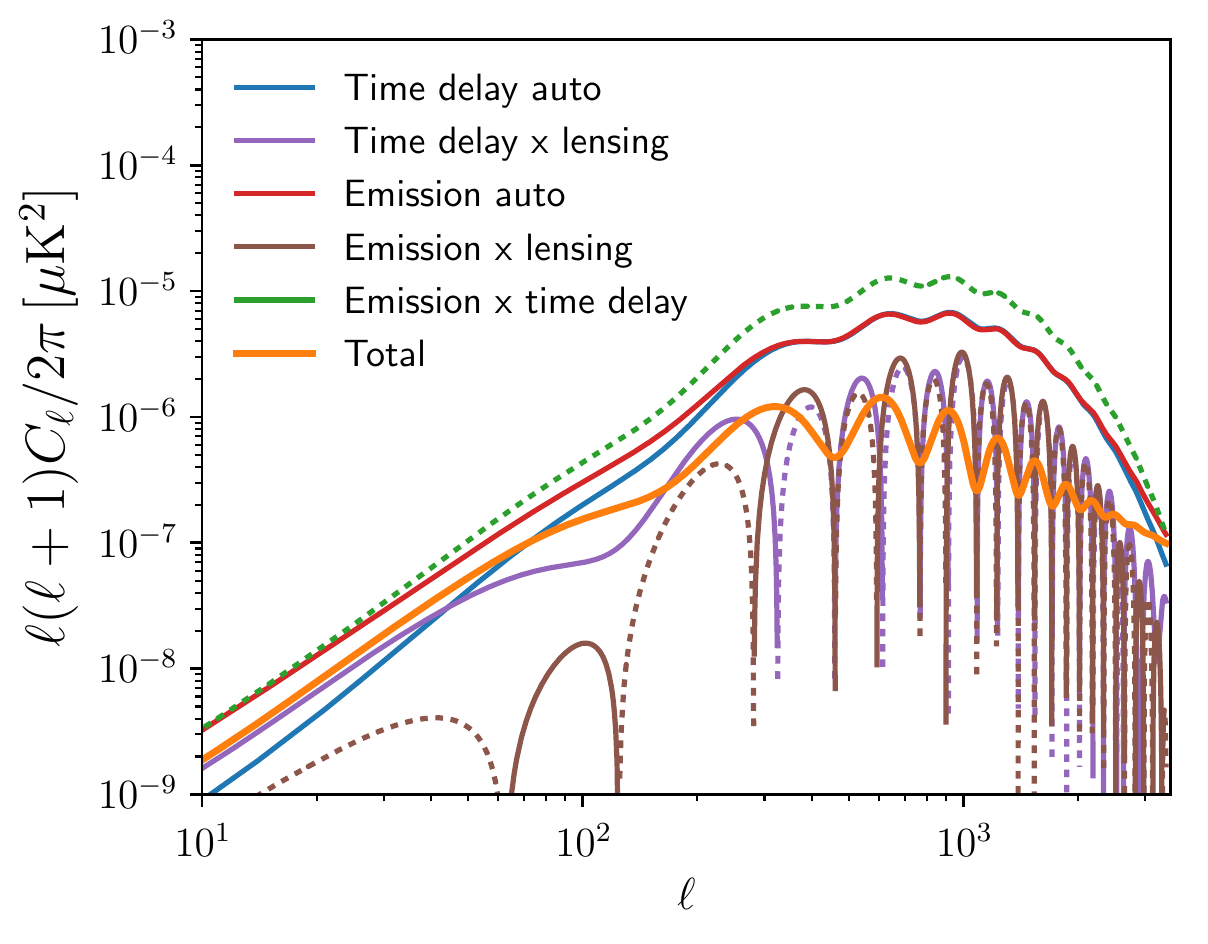}
\caption{
Contributions to the correction to the standard lensed B-mode power spectrum from
emission angle and time delay effects, including their cross correlation and cross-correlation with standard lensing.
The time delay contributions are consistent with Ref.~\cite{Hu:2001yq}, but
the emission angle and time delay terms are anti-correlated (dotted lines are negative) and the total correction (orange) has substantial cancellations between terms.
\label{BBterms}
}
\end{figure}

Expanding the polarization source tensor at the perturbed location $\calS_{ab}([\chi_*+\delta\chi]\vnhat',\eta_*)$
to linear order in $\delta\chi$,
the time delay gives the second-order correction  to the observed polarization
\begin{equation}
\Delta \tilde{P}_{ab}^{(2)}|_{\rm delay} =  \delta \chi \frac{\ud P_{ab}}{\ud\chi_*} =
\frac{\delta\chi}{\chi_*} \grad_{\la a}\grad_{b\ra} \psi_t,
\end{equation}
where $\delta\chi/\chi_* = \psi_d$ and we defined the polarization derivative potential\footnote{The \CAMB\ source for $\psi_t$ is $\frac{\chi_*}{4k^2\chi^2}\frac{\ud}{\ud \eta}(g \zeta)$ where $g$ is the visibility prior to reionization and the sign is using \CAMB's convention.}
$\psi_t = \chi_\ast \ud \psi_E / \ud \chi_\ast$:
\begin{eqnarray}
\psi_t &\equiv&
-\frac{\chi_*}{4}e^{-\tau}\int \frac{\ud^3 \vk}{(2\pi)^3} \zeta(\vk) \frac{\ud}{\ud \chi_*} \left(\frac{e^{i\vk\cdot\vnhat\chi_*}}{k^2\chi_*^2}\right).
\label{psit}
\end{eqnarray}
The time-delay contribution to the second-order B-mode polarization is then
\begin{equation}
\Delta B(\vell)|_{\rm delay} = \int \frac{\ud^2\vL}{(2\pi)^2} L^2 \sin(2\varphi_{L\ell}) \psi_t(\vL) \psi_d(\vell-\vL).
\end{equation}
The time-delay auto-spectrum together with the cross-correlation with the lensing and emission-angle terms gives an additional contribution to the power spectrum
\begin{multline}
\Delta \tilde C^{BB}_\ell|_{\rm delay} = \int \frac{\ud^2\vL}{(2\pi)^2}\biggl(
L^4 \sin^2(2\varphi_{L\ell}) C^{\psi_{t}}_L C^{\psi_d}_{|\vell-\vL|}
\\
+4L^2 \sin(2\varphi_{L\ell})[\vell\times \vL - L^2 \sin(2\varphi_{L\ell})] C^{\psi_{\zeta}\psi_t}_L C^{\psi_d}_{|\vell-\vL|}
\\
-2L^2 \sin^2(2\varphi_{L\ell})\vL\cdot(\vell-\vL)C^{E\psi_{t}}_L C^{\phi\psi_d}_{|\vell-\vL|}
\biggr).
\label{CBtimeemission}
\end{multline}
The first and last term have been calculated before in Ref.~\cite{Hu:2001yq}, but the middle term is the new contribution from the correlation with the emission-angle term.
The time delay and emission angle B-modes are anti-correlated and have significant cancellations; see Fig.~\ref{BBterms}.

To understand these cancellations, we can use Eq.~\eqref{psizeta} to write Eq.~\eqref{psit} as
\begin{eqnarray}
\psi_t
&=& \psi_\zeta+\frac{e^{-\tau}}{4}\int \frac{\ud^3\vk}{(2\pi)^3}\frac{\zeta(\vk)}{k^2\chi_*^2} e^{i\vk\cdot\vnhat\chi_*}\nonumber \\
&=&
\psi_\zeta - \psi_E.
\label{psitE}
\end{eqnarray}
Using this we can then also write the second-order time-delay polarization as
\begin{eqnarray}
\Delta \tilde{P}_{ab}^{(2)}|_{\rm delay} &=&  \psi_d
\left(
\grad_{\la a}\grad_{b\ra} \psi_\zeta - P_{ab}\right) \nonumber\\
&=&\grad_{\la a}\grad_{b\ra}(\psi_d \psi_\zeta)
- 2\grad_{\la a} \psi_d \grad_{b\ra} \psi_\zeta
\nonumber\\&&\qquad
- \psi_\zeta\grad_{\la a}\grad_{b\ra}\psi_d  - \psi_d  P_{ab}.
\label{Pol_ab_time}
\end{eqnarray}
Note that, as shown in Fig.~\ref{poldeltacl}, due to the extra derivatives in $\psi_\zeta$ and $\psi_t$, we have $C_\ell^{\psi_E} = C_\ell^{E}/\ell^4 \sim C_{\ell}^{\psi_\zeta}/\ell^2$, so $\psi_\zeta\sim\psi_t$ on relevant scales, though both are typically smaller than $\ell\psi_{E}$ because radial derivatives are geometrically suppressed in the angular power~\cite{Hu:2001yq}.
The first term in Eq.~\eqref{Pol_ab_time}, $\grad_{\la a}\grad_{b\ra}(\psi_d \psi_\zeta) $, is close to $\grad_{\la a}\grad_{b\ra}(\psi_d \psi_t) $, which represents the change in the standard transverse polarization components due to evaluating the polarization potential $\psi_E$ at the radially-displaced emission location.
The time delay itself has a large coherence length (little small-scale power) and is very small, with RMS amplitude $\la \left(\delta \chi/\chi_*\right)^2\ra^{1/2} = \la\psi_d^2\ra^{1/2} \approx 5\times 10^{-5}$ (corresponding to a comoving distance of about $0.7\,\Mpc$). This term is therefore negligibly small,
and is highly suppressed on large scales as it is a total derivative. It is also pure E-mode, and hence does not contribute to the B-mode power.
The second term $- 2\grad_{\la a} \psi_d \grad_{b\ra} \psi_\zeta$ is minus the second-order polarization from the emission-angle effect, and hence cancels with it.
The $-\psi_\zeta\grad_{\la a}\grad_{b\ra}\psi_d $ term depends on derivatives of the emission angle, and describes the contribution of new components of the polarization tensor due to anisotropic curvature of the last-scattering surface. The final term $\psi_d  P_{ab}$
\AL{encodes the effect on the standard polarization of the perturbed angular-diameter distance to the perturbed last-scattering surface, and }is important because it correlates with the substantially larger second-order lensing term $\valpha\cdot \vgrad P_{ab}$.
\AC{What physical effect is being described in the previous sentence?}
\AL{I'm trying to split the change in the polarization into a term involving the change in the source viewed at the background distance, and the background source viewed at the perturbed distance, where the latter effect is the one that correlates well with lensing (corresponding to a shear-free dilation).
I would agree this could be on firmer footing and scope for clarification... e.g. really it's the effect of the angular diameter distance relative to $S^{1}$.
 }

Combining the second-order time-delay contribution of Eq.~\eqref{Pol_ab_time} with the emission-angle term the total can then be written as
\begin{eqnarray}
\Delta \tilde{P}_{ab}^{(2)}|_{\rm tot} &=&  \psi_d\left(
\grad_{\la a}\grad_{b\ra} \psi_\zeta - P_{ab}\right) + 2\grad_{\la a} \psi_d \grad_{b\ra} \psi_\zeta
\nonumber\\
&=&
\grad_{\la a}\grad_{b\ra}(\psi_d \psi_\zeta) - \psi_\zeta\grad_{\la a}\grad_{b\ra}\psi_d
 - \psi_d  P_{ab},
\nonumber\\
\label{totdeltaPab}
\end{eqnarray}
where the second term in the final line dominates the auto-correlation and the last term dominates the correlation with lensing deflection.
Note that since $2\grad_{\la a} \psi_d \grad_{b\ra} \psi_\zeta$ has cancelled, there are no remaining significant $\psi_\zeta$ terms if the last-scattering surface is locally at a constant angle to the background one ($\vd=\vgrad\psi_d$ approximately constant on the scale of the radiation perturbations). The dominant contribution now only depends on the curvature of the perturbed last-scattering surface.
Since the power spectrum of $\psi_d$ peaks on large scales, this substantially reduces the total signal compared to either time delay or emission angle individually, explaining the cancellations seen in Fig.~\ref{BBterms} and the different scale dependence of the total.
The fact that the perturbed last-scattering surface remains normal to the emitted direction means that the dominant correction to the small-scale polarization from a large-scale lens is
obtained by applying the standard
differential operators to the radially-displaced potential (approximately $\grad_{\la a}\grad_{b\ra}(\psi_d \psi_\zeta)$, which is pure E), ensuring that the production of B modes is suppressed. The bulk of the second-order E-mode signal from time delay is also cancelled in the power spectrum because radial displacements of the source plane are nearly statistically equivalent by statistical homogeneity~\cite{Hu:2001yq}.

\begin{figure}
\includegraphics[width = 0.45\textwidth]{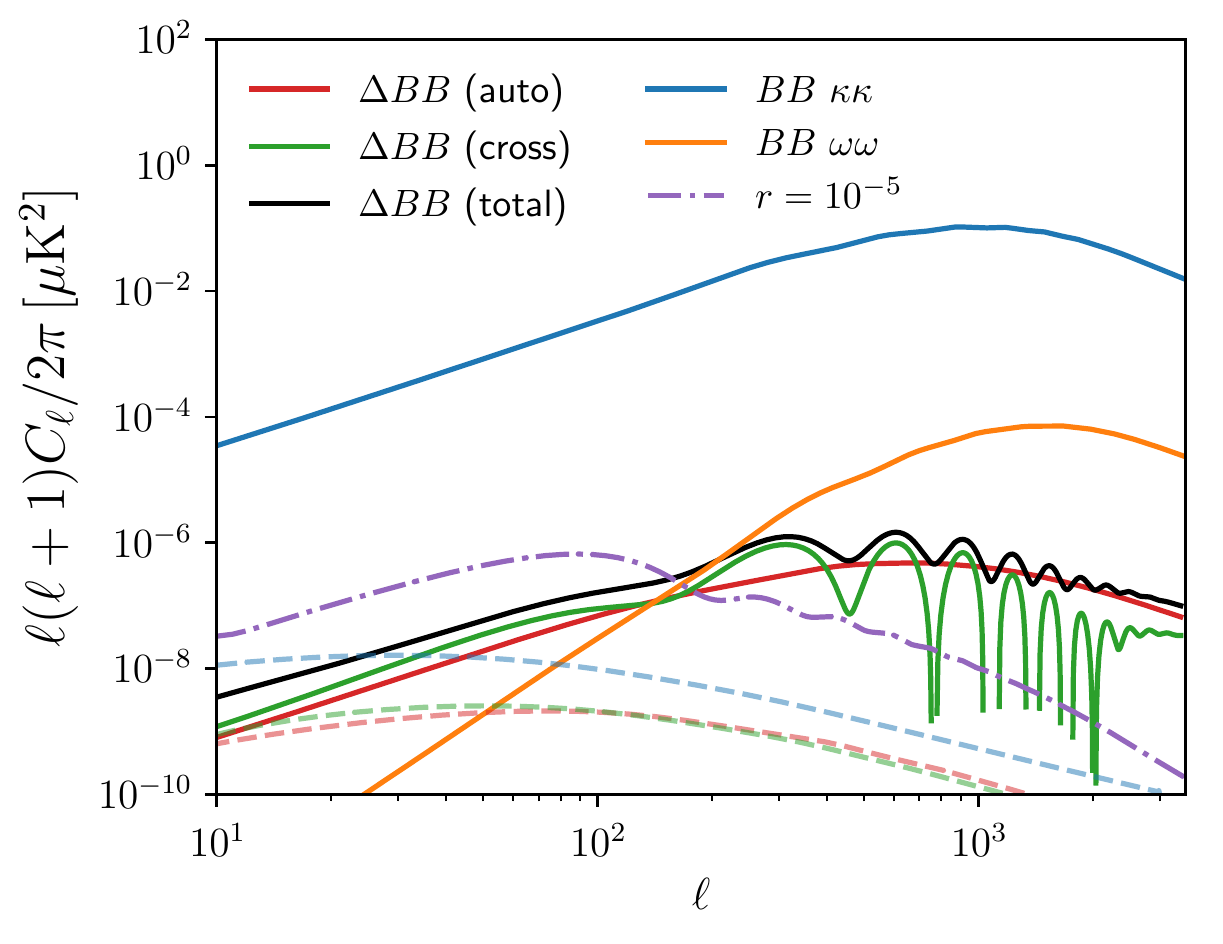}
\caption{
Corrections to the standard lensed B-mode power spectrum from the total of time-delay and emission-angle effects, compared to standard gradient (blue) and post-Born curl (orange) lensing signals. The green line shows the cross-correlation with standard lensing, which is an important part of the signal on large scales but could be substantially removed by standard delensing.
On the scales most relevant for primordial gravitational waves, the total is larger than or comparable to the B modes from post-Born curl lensing. The auto spectrum of the new terms is approximately white on large scales, and dominates curl lensing at $\ell \alt 100$.
The dashed lines show the approximate contribution from B-mode power generated by lensing of polarization from reionization to the similarly coloured solid lines.
For reference the dash-dotted line shows the B-mode power expected from primordial gravitational waves with tensor-to-scalar ratio $r=10^{-5}$.
Additional bispectrum contributions~\cite{Pratten:2016dsm} are not shown and would be partly removed by standard delensing.
\label{totalBB}
}
\end{figure}

Expanding into harmonics the total second-order contribution to the B modes including time delay, emission angle and lensing gives
\begin{multline}
\Delta B(\vell) = -\int \frac{\ud^2\vL}{(2\pi)^2}\biggl(
 L^2 \sin(2\varphi_{L\ell})\psi_{\zeta}(\vell-\vL) \psi_d(\vL)
 \\
+\sin(2\varphi_{L\ell}) E(\vL)[\psi_d(\vell-\vL)
+ \vL\cdot(\vell-\vL)\phi(\vell-\vL)]
\biggr),
\label{Bdeltatot}
\end{multline}
where, as before, $E$ is the unlensed E-mode polarization excluding reionization.
The corresponding power spectrum is
\begin{multline}
\Delta \tilde C^{BB}_\ell \approx \int \frac{\ud^2\vL}{(2\pi)^2}\biggl( 
 L^4\sin^2(2\varphi_{L\ell}) C^{\psi_{\zeta}}_{|\vell-\vL|} C^{\psi_d}_L
\\-4\left[\vell\times \vL - \frac{L^2}{2} \sin(2\varphi_{L\ell})\right]\sin (2\varphi_{L\ell})C^{E\psi_{\zeta}}_L\\
\qquad\qquad\qquad\times \left[C^{\psi_d}_{|\vell-\vL|}+  \vL\cdot(\vell-\vL)C^{\phi\psi_d}_{|\vell-\vL|} \right]
\\
+\sin^2(2\varphi_{L\ell}) C^{EE}_L \left[C^{\psi_d}_{|\vell-\vL|}+2\vL\cdot(\vell-\vL) C^{\phi\psi_d}_{|\vell-\vL|} \right]
\biggr).
\label{CBtotapprox}
\end{multline}
The middle term involving $C_L^{E\psi_\zeta}$
is a small part of the total except on small scales because of the low correlation between $E$ and $\psi_\zeta$. The $C_{|\vell-\vL|}^{\psi_d}$ parts of the second and third term are negligible compared to the lensing cross-correlation terms (which are enhanced by derivatives).
The dominant corrections to the lensed B-mode power spectrum are therefore from the auto-spectrum of the first term in Eq.~\eqref{Bdeltatot} and the cross-correlation of the second and third (lensing) terms there.
Numerical results are shown in  Fig.~\ref{totalBB}.

On large scales, with contributions dominated by $L \gg \ell$, we have an approximately white-noise spectrum with
\begin{eqnarray}
\Delta \tilde C^{BB}_\ell &\approx& \int \frac{ L^3\ud L}{4\pi}\biggl(
L^2 C^{\psi_{\zeta}}_L C^{\psi_d}_L
- 2 C^{EE}_L C^{\phi\psi_d}_L\biggr) \nonumber\\
&\approx& 1.2 \times 10^{-10}\, \mu {\rm K}^2.
\label{BBtotalwhite}
\end{eqnarray}
About two-thirds of the total comes from the second term involving the time-delay correlation with lensing.
The first (auto-spectrum) term is only a quarter the size of the emission-angle term (Eq.~\ref{BBemissionWhite}) due
to cancellation with time delay, and corresponds to an effective noise of approximately
$0.02,\mu {\rm K}\,\arcmin$ that would not be removed by standard delensing. Equation~\eqref{BBtotalwhite} is accurate at the $10\,\%$ level for $\ell \alt 80$, but overestimates the total by around $20\,\%$ by $\ell \approx 100$ as the cross-correlation term starts to be significantly $\ell$ dependent.
On large scales there is also a contribution from lensing of polarization generated at reionization. We estimate this using the same results as for the recombination signal\footnote{We use the full-sky result of Appendix~\ref{FullSky} for numerical results, though the flat-sky result agrees quite well even with such a large-scale source.}, but using the reionization visibility and approximating the redshift of the reionization source as $z\approx 8.1$. This signal is not white, and dashed lines in Fig.~\ref{totalBB} show that, though small, the contribution to the time delay and emission signal is not negligible at $\ell \ll 100$ (comparable at $\ell\sim 10$). The new contributions are relatively more important for the reionization signal because the standard deflection signal (blue dashed in the figure) is not as derivatively enhanced as the recombination signal due to the absence of significant small-scale unlensed E-mode power from reionization.

The large-scale cancellation between time-delay and emission-angle effects can also be seen directly from the total correction to the polarization tensor from these effects by writing Eq.~\eqref{totdeltaPab} as
\begin{eqnarray}
\Delta \tilde{P}_{ab}^{(2)}|_{\rm tot} &=&
\grad_{\la a}\bigl( \psi_d \grad_{b\ra} \psi_\zeta\bigr) + \grad_{\la a} \psi_d \grad_{b\ra} \psi_\zeta
- \psi_d  P_{ab}.
\nonumber\\
\label{totdeltaPabTwo}
\end{eqnarray}
The first term on the right is a total derivative, and gives a blue rather than white power spectrum on large scales (and hence is relatively suppressed there). The second term is half of the emission-angle contribution, so the total large-scale auto-spectrum power from that term is one-quarter that from the emission angle alone.

On small scales ($\ell\gg 1000$) where $C_\ell^{\psi_\zeta}$ is small
the auto-spectrum of the first term in Eq.~\eqref{Bdeltatot}
follows the emission angle power:
\begin{equation}\label{BBtotsmall}
  \Delta \tilde C^{BB}_\ell|_{\rm auto} \approx 4\ell^2 C_\ell^{\psi_d}
  \int \frac{\ud \log L}{4\pi} L^4 C_L^{\psi_\zeta}.
\end{equation}
However, on small scales this (and the total) is small compared to the B-mode power produced by field rotation (i.e., lensing by post-Born curl modes).
Note that the equivalent result for the auto-power spectrum of time delay alone (the first term on the right of Eq.~\ref{CBtimeemission}) couples to the power spectrum of the time delay ($\psi_d$) directly rather than its gradient.



\section{Conclusions}

Second-order effects can create important signals in the CMB, and are dominated by CMB lensing.
We calculated new predictions for the effect of the photon emission angle not being orthogonal to the background last-scattering surface, which is not accounted for in the standard lens- remapping approximation. The emission angle is enhanced to the $\clo(\alpha)\sim 10^{-3}$ level by the large distance to last scattering, but is not derivatively coupled like the main lensing-displacement effect and hence still very small. For the temperature and E-mode polarization, the correction to the power spectrum is dominated by the correlation with the larger lensing-displacement correction, but is negligible compared to cosmic variance. For the B-mode polarization, the spectrum is about $10^{-4}$ of the dominant lensing B-mode spectrum. It is partly cancelled by the contribution from time delay, but the total remains substantially larger than locally-sourced second-order effects at recombination  on the scale of the recombination peak of the tensor B-mode power spectrum~\cite{Mollerach:2003nq,Fidler:2014oda}.
Since the dominant B-mode signal can be largely removed by delensing, the total emission angle and time delay signal could ultimately act as an important extra source of noise confusion for primordial gravitational waves if the tensor-to-scalar ratio $r\ll 10^{-5}$ (if unaccounted for it would contribute a bias in the range $10^{-6} \alt \Delta r \alt 2\times 10^{-6}$ for tensor B-mode measurements at $\ell \agt 30$ with substantial delensing).

At next order in deflections, post-Born lensing can lead to a more complicated lensing geometry and generate polarization rotation. While polarization rotation can efficiently convert E-mode into B-mode polarization, the rotation angle is only $\clo(\alpha^2)$ (sub-arcsecond) and hence negligible compared to other signals: polarization rotation can safely be neglected, as it has been in most previous post-Born calculations~\cite{Hirata:2003ka,Pratten:2016dsm}. Post-Born lensing also generates field-rotation (curl deflection) sourced B-modes, which dominate the emission angle B-modes on small scales because the field rotation is enhanced by an additional derivative of the unlensed polarization. However, the post-Born curl B modes are total derivatives so their spectrum is suppressed by a factor of $\ell^2$ on large scales. The combination of the emission-angle and time-delay signals has a white spectrum on large angular scales, and therefore dominates on the scales relevant for searches for B modes sourced by primordial gravitational wave. This may be the leading signal that cannot in principle be delensed.

The code to produce our numerical results is available on GitHub\footnote{ \url{https://github.com/cmbant/notebooks/blob/master/EmissionAngle.ipynb}} and the final result for the total B-mode power can be calculated easily using the module in Python \CAMB\footnote{\url{http://camb.readthedocs.io/en/latest/emission_angle.html}}.

\quad

\section*{Acknowledgments}
We thank Julian Carron, Giulio Fabbian, Christian Fidler and Geraint Pratten for discussion.
\ALorcid\ acknowledges support from the European Research Council under
the European Union's Seventh Framework Programme (FP/2007-2013) / ERC Grant Agreement No. [616170],
and \ALorcid\ and AC from the Science and Technology Facilities Council {[}grant numbers ST/L000652/1 and ST/N000927/1, respectively{]}. AH is supported by a United Kingdom Space Agency \emph{Euclid} grant, and an STFC Consolidated Grant.


\appendix
\begin{widetext}

\section{Emission-angle corrections to the lensed temperature from the quadrupole source}
\label{app:temppol}

There is an additional contribution to the CMB temperature anisotropy from the radiation quadrupole (temperature and E-mode polarization) at recombination, but this is not a large effect.
The relevant quadrupole source tensor is the same as the source of linear polarization,
$\calS_{ab} \equiv -\zeta_{ab}/4$, which is discussed in Sec.~\ref{sec:polarization}. The corresponding contribution to the lensed temperature observed in direction $\vnhat$ is given by the source at $\vx = \chi_*\vnhat'$ projected along the emission direction $\ve$:
\begin{align}
\tilde{T}|_{\rm quad}(\vnhat) &= -e^{-\tau}\calS_{ab}(\chi_\ast \vnhat') e^ae^b \\
&= -e^{-\tau}\left(
\cos^2 \! d\, \hat{n}^{\prime a} \hat{n}^{\prime b} - \sin(2d) \hat{d}^{\prime a} \hat{n}^{\prime b} + \sin^2 \! d\, \hat{d}^{\prime a} \hat{d}^{\prime b}\right) \calS_{ab} \\
&= -e^{-\tau} \left[\left(1-\frac{3}{2}d^2\right) \calS_{ab}(\chi_\ast \vnhat') \hat{n}^{\prime a} \hat{n}^{\prime b} - 2 d^{\prime a} \calS_a^{(1)}(\vnhat';\chi_\ast \vnhat') + [\calS_{ab}]^{\rm TT}(\vnhat';\chi_\ast \vnhat') d^{\prime a} d^{\prime b}+ \cdots \right]
\end{align}
correct to third order. Here, the final term involves the symmetric, trace-free projection (TT) of the source tensor orthogonal to $\vnhat'$. The second-order correction to $\tilde{T}|_{\rm quad}(\vnhat)$ from the emission-angle effect is simply
$\Delta \tilde{T}^{(2)}|_{\rm quad} = 2 e^{-\tau} d^a \calS_a^{(1)}(\vnhat;\chi_\ast \vnhat)$, which can be
written as $\Delta \tilde{T}^{(2)}|_{\rm quad}  = 2\vd\cdot \vgrad\psi_\zeta$ following the definitions in Sec.~\ref{sec:polemission}. For the third-order corrections, we have
\begin{equation}
\Delta \tilde{T}^{(3)}|_{\rm quad} = -\frac{3}{2} d^2 T_Q - d^a d^b P_{ab} +2 \vd\cdot(\valpha\cdot \vgrad \vgrad\psi_\zeta),
\end{equation}
where all terms are evaluated in direction $\vnhat$. Here,
$T_Q(\vnhat) = -e^{-\tau} \calS_{ab}(\chi_\ast \vnhat) \hat{n}^a \hat{n}^b$
is the quadrupole-sourced part of the unlensed temperature and $P_{ab}(\vnhat) =  e^{-\tau} [\calS_{ab}]^{\rm TT}(\vnhat;\chi_\ast \vnhat)$ is the unlensed polarization tensor.
Expanding in harmonics and calculating the power spectrum (including the cross-correlation with lensing, $\valpha \cdot \vgrad T$, and the Doppler term, $\vd\cdot \vgrad\psi_v$) gives
\begin{multline}
\Delta \tilde C_\ell^{TT}|_{\rm quad} =
2\int \frac{\ud^2 \vL}{(2\pi)^2}  [\vL\cdot(\vell-\vL)]^2
C_L^{\phi\psi_d} C_{|\vell-\vL|}^{T\psi_\zeta}
+2\int \frac{\ud^2 \vL}{(2\pi)^2}  [\vL\cdot(\vell-\vL)]^2
C_L^{\psi_d} \left[C_{|\vell-\vL|}^{\psi_v\psi_\zeta}+ 2 C_{|\vell-\vL|}^{\psi_\zeta}\right]
\\
-2 \ell^2 C^{T\psi_\zeta}_{\ell}\int \frac{\ud L}{L} \frac{L^4 C^{\phi\psi_d}}{2\pi}
-3C_\ell^{TT_Q}\int \frac{\ud L}{L} \frac{L^4 C^{\psi_d}_L}{2\pi} .
\end{multline}
The first and third terms, those involving lensing, are largest but partly cancel. All terms have
$|\ell^2\Delta\tilde C^{TT}_\ell/2\pi| \alt 0.04\,\muK^2$, so they are all safely negligible compared to cosmic variance (as expected).

\section{CMB power spectra from field and polarization rotation}
\label{app:rotpower}
We calculate the effect of lensing on power spectra of the CMB temperature (T), and polarization (E and B) fields.
The post-Born lensing deflection angle may be decomposed into two lensing potentials
\begin{align}
\alpha_a &= \nabla_a \phi + \epsilon_{ab} \nabla^b \Omega ,
\end{align}
\n
where $\phi$ is the lensing potential (describing convergence $\kappa = -\vgrad^2\phi/2$), $\Omega$ is the curl potential (describing rotation $\omega = -\vgrad^2\Omega/2$), and $\nabla$ is the angular covariant derivative.

Working in the flat-sky approximation, the lensed temperature anisotropies can be written as
\begin{align}
\tilde{T} (\vnhat) &= \int \frac{\ud^2 \vell}{(2 \pi)^2} \tilde{T}({\vell}) \, e^{i \vell \cdot \vnhat} .
\end{align}
\n
The lensed polarization anisotropies can be defined in an analogous way in terms of the spin-$\pm2$ Stokes parameters $\tilde{Q} \pm i \tilde{U}$ and the more physically relevant $\tilde{E}$ and $\tilde{B}$ modes
\begin{align}
\left(\tilde{Q} \pm i \tilde{U} \right) (\vnhat) &= - \int \frac{\ud^2 \vell}{(2\pi)^2} \, \left( \tilde{E}({\vell}) \pm i \tilde{B}({\vell}) \right) \, e^{\pm i 2 \varphi_{\vell}} \, e^{i \vell \cdot \vnhat} .
\end{align}
\n
Here, $\varphi_\vell$ is the angle that $\vell$ makes with the $x$-direction.
The unlensed CMB observables $\lbrace T(\vell),E(\vell),B(\vell) \rbrace$ are defined in exactly the same way. The lensed (or unlensed) angular power spectra in the flat-sky approximation are then
\begin{align}
\la \tilde{X}({\vell})\tilde{Y}({\vell'}) \ra &= (2 \pi)^2 \delta_D(\vell + \vell') \, \tilde{C}^{XY}_{\ell} .
\end{align}
Neglecting emission angle effects, we expand a lensed spin-s field $\tilde{\zeta}^s(\vnhat)$ as a perturbative expansion in terms of the unlensed field $\zeta^{s}$ as \cite{Hu:2000ee,Lewis:2006fu}
\begin{align}
\tilde{\zeta}^s (\vnhat) &= e^{-is\beta}\left(\zeta^s (\vnhat) + \alpha_a \nabla^a \, \zeta^s (\vnhat) + \frac{1}{2} \alpha_a \alpha_b \nabla^a \nabla^b \, \zeta^s (\vnhat) + \frac{1}{6} \alpha_a \alpha_b \alpha_c \nabla^a \nabla^b \nabla^c \, \zeta^s (\vnhat) + \mathcal{O}(\alpha^4)\right).
\label{eq:alphaseries}
\end{align}
Here the $e^{-is\beta}$ prefactor accounts for the rotation compared to geodesic parallel transport between $\vnhat'$ and $\vnhat$ as described in Sec.~\ref{sec:polarization}.

Expanding in harmonics we have the leading terms
\begin{multline}
\tilde{\zeta}^s(\vell) \approx  {\zeta}^s(\vell) - \int \frac{d^2 \vell_1}{(2 \pi)^2} \, \zeta^s (\vell_1) \, e^{i s \varphi_{\vell_1 \vell}} \, \left[ \vell_1 \cdot \left( \vell - \vell_1 \right) \phi(\vell - \vell_1) + \vell_1 \times \vell \,\Omega (\vell - \vell_1) + i s \,\beta(\vell-\vell_1) \right] \\
 -\frac{1}{2} \int \frac{d^2 \vell_1}{(2 \pi)^2} \int \frac{d^2 \vell_2}{(2 \pi)^2} \, \zeta^s (\vell_1) \, e^{i s \varphi_{\vell_1 \vell}}  \,\biggl( \left[ \vell_1 \cdot \vell_2 \, \phi(\vell_2) + \vell_1 \times \vell_2 \, \Omega(\vell_2) \right] \\
\hfill\times \left[ \vell_1 \cdot \left( \vell_2 + \vell_1 - \vell \right) \, \phi(\vell - \vell_1 - \vell_2) + \vell_1 \times (\vell_2 - \vell) \, \Omega(\vell - \vell_1 - \vell_2) \right]\qquad
\\
+ s^2 \beta(\vell_2) \beta( \vell-\vell_1 - \vell_2)
-2is    \left[ \vell_1 \cdot \vell_2 \, \phi(\vell_2) + \vell_1 \times \vell_2 \, \Omega(\vell_2) \right] \beta(\vell-\vell_1-\vell_2)
\biggr) + \cdots,
\end{multline}
where the cross product is defined by $\vell \times \vL \equiv \epsilon_{ab} \ell^a L^b$ and $\varphi_{\vell_1 \vell} = \varphi_{\vell_1} - \varphi_{\vell}$ is the angle between $\vell_1$ and $\vell$.
If we assume zero unlensed B modes, we can then relate this to the expansion of the $X\in{T,E,B}$ fields as follows:
\begin{multline}
\tilde{X}(\vell) \approx  X(\vell) - \int \frac{\ud^2 \vell_1}{(2 \pi)^2} \, \bar{X} (\vell_1) \, \left\{\trig_X(s\varphi_{\vell_1 \vell}) \, \left[ \vell_1 \cdot \left( \vell - \vell_1 \right) \phi(\vell - \vell_1) + \vell_1 \times \vell \,\Omega (\vell - \vell_1) \right] - s \, \trigbar_X(s\varphi_{\vell_1 \vell}) \,\beta(\vell-\vell_1) \right\}\\
 -\frac{1}{2} \int \frac{\ud^2 \vell_1}{(2 \pi)^2} \int \frac{\ud^2 \vell_2}{(2 \pi)^2} \, \bar{X}(\vell_1) \biggl(  \trig_X(s \varphi_{\vell_1 \vell})  \,\bigl\{ \left[ \vell_1 \cdot \vell_2 \, \phi(\vell_2) + \vell_1 \times \vell_2 \, \Omega(\vell_2)
  \right] \\
\hfill\times \left[ \vell_1 \cdot \left( \vell_2 + \vell_1 - \vell \right) \, \phi(\vell - \vell_1 - \vell_2) + \vell_1 \times (\vell_2 - \vell) \, \Omega(\vell - \vell_1 - \vell_2)\right] + s^2 \beta(\vell_2)\beta(\vell-\vell_1-\vell_2) \bigr\}\qquad
\\
+2s\,  \trigbar_X(s \varphi_{\vell_1 \vell})  \left[ \vell_1 \cdot \vell_2 \, \phi(\vell_2) + \vell_1 \times \vell_2 \, \Omega(\vell_2) \right] \beta(\vell-\vell_1-\vell_2)
\biggr) + \cdots,
\label{eqn:expansion}
\end{multline}
where $s=2$ for $X\in{E,B}$ and $s=0$ for $X=T$, $\trig_E = \cos$, $\trig_B=\sin$, $\trigbar_E=\sin$, $\trigbar_B=-\cos$, $\trig_T=1$, $\trigbar_T=0$,
and $\bar{T}=T$, $\bar{E}=E$ and $\bar{B}=E$.
For example the leading-order B mode from image and polarization rotation is
\begin{equation}
  \tilde B(\vell) \supset \int \frac{\ud^2 \vell'}{(2 \pi)^2} \, E(\vell')
\left[ \sin(2\varphi_{\vell'\vell})\, \vell \times \vell'\Omega(\vell-\vell')
  -2\cos(2\varphi_{\vell'\vell})\,\beta(\vell-\vell')\right]
,
\end{equation}
where the first term is from curl deflections and the second from polarization rotation.

To leading order for Gaussian fields, the lensed CMB power spectra contributions from curl and rotation are then\footnote{This curl result differs from Eq.~(52) of Ref.~\cite{Hirata:2003ka}, which uses an incorrect definition of the angle $\alpha$ in their trigonometric factors, giving numerical results for the $\tilde C_\ell^{BB}$ that are too large on small scales. References~\cite{Marozzi:2016und,Marozzi:2016qxl} incorrectly set $\beta=\omega$, giving numerical results that are wrong by orders of magnitude on both large and small scales.}
~\cite{Hirata:2003ka,Cooray:2005hm,Marozzi:2016qxl}

\begin{eqnarray}
\Delta\tilde{C}_\ell^{BB} &=&  \int \frac{\ud^2 \vell_1}{(2 \pi)^2}  C_{\ell_1}^{EE} \left(
\sin^2(2\varphi_{\vell_1 \vell}) (\vell\times \vell_1)^2 C^\Omega_{|\vell-\vell_1|}
-2\sin(4\varphi_{\vell_1 \vell})\vell\times \vell_1 C^{\Omega\beta}_{|\vell-\vell_1|}
+ 4\cos^2(2\varphi_{\vell_1\vell}) C^\beta_{|\vell-\vell_1|} \right) \\
\Delta\tilde{C}_\ell^{EE} &=&
\left(-\ell^2R^\Omega  - 4 \la\beta^2\ra\right) C_\ell^{EE}
\nonumber\\&& \hfill+ \int \frac{\ud^2 \vell_1}{(2 \pi)^2}  C_{\ell_1}^{EE}\left(
\cos^2(2\varphi_{\vell_1 \vell})(\vell\times \vell_1)^2C^\Omega_{|\vell-\vell_1|}
+2\sin(4\varphi_{\vell_1 \vell})\vell\times \vell_1 C^{\Omega\beta}_{|\vell-\vell_1|}
+ 4\sin^2(2\varphi_{\vell_1\vell}) C^{\beta}_{|\vell-\vell_1|}  \right) \\
\Delta\tilde{C}_\ell^{TE} &=&
 \left(-\ell^2R^\Omega  - 2 \la\beta^2\ra\right) C_\ell^{TE}
\nonumber\\&&\hfill+ \int \frac{\ud^2 \vell_1}{(2 \pi)^2}  C_{\ell_1}^{TE}\left(
\cos(2\varphi_{\vell_1 \vell})\vell\times \vell_1 C^\Omega_{|\vell-\vell_1|}+
2\sin(2\varphi_{\vell_1\vell}) C^{\beta\Omega}_{|\vell-\vell_1|} \right)\vell\times \vell_1\\
\Delta\tilde{C}_\ell^{TT} &=&
 \left(-\ell^2R^\Omega \right) C_\ell^{TT}
 + \int \frac{\ud^2 \vell_1}{(2 \pi)^2}  C_{\ell_1}^{TT}C^\Omega_{|\vell-\vell_1|}\left(\vell_1\times \vell\right)^2,
\end{eqnarray}
where
\begin{equation}
R^\Omega \equiv \int \frac{ \ud \log \ell}{4\pi} \ell^4 C_\ell^\Omega,
\qquad\qquad \la \beta^2\ra = \int \ud \log \ell \,\frac{\ell^2 C_\ell^\beta}{2\pi}.
\end{equation}




\section{Full-sky lensed B-mode calculation}
\label{FullSky}

On large angular scales the flat-sky calculation presented in the main text is not expected to be accurate. In this appendix we give the full-sky calculation of the B-modes induced by emission-angle and time-delay effects. We ignore post-Born effects here as they are subdominant on the scales relevant for full-sky corrections.

As shown in Sec.~\ref{sec:delay}, the perturbation to the polarization tensor due to the perturbed emission angle and time delay is $\Delta \tilde P_{ab}^{(2)}  = \grad_{\la a} \grad_{b \ra}(\psi_d \psi_\zeta) - \psi_\zeta \grad_{\la a} \grad_{b \ra} \psi_d - \psi_d P_{ab}$. The first term is pure E mode, and does not contribute to the B mode. The perturbed complex polarization is $\Delta
{}_{\pm 2} \tilde{P} = e^a_{\pm} e^b_{\pm}\Delta \tilde P_{ab}^{(2)} $ where $e^a_{\pm} = \hat{\theta}^a \pm i \hat{\phi}^a$ are null basis vectors on the sphere. The complex polarization ${}_{\pm 2}P$ is spin-$\pm2$ on this basis, with the line-of-sight direction $\hat{n}^a$ completing an orthonormal right-handed set\footnote{Note that this differs from the flat-sky convention used in the main text.} $\{ \hat{\theta}^a, \hat{\phi}^a, \hat{n}^a\}$. Expanding the potentials in spherical harmonics and noting that $e^a_{\pm} e^b_{\pm} \grad_a \grad_b Y_{lm} = \sqrt{(l+2)!/(l-2)!}{}_{\pm2}Y_{lm}$ gives
\begin{equation}
\Delta {}_{\pm 2}\tilde{P} =  e^a_{\pm} e^b_{\pm} \grad_a \grad_b (\psi_d \psi_\zeta)-\sum_{L'M'} \sum_{LM} \left[ \psi_{\zeta,LM} \psi_{d,L'M'} \sqrt{(L'+2)!/(L'-2)!} + \psi_{d,LM} E_{L'M'} \right] {}_{\pm2}Y_{L'M'} Y_{LM}.
\label{QUfullsky}
\end{equation}
Using the harmonic expansion definition
\begin{equation}
{}_{\pm 2} P(\vnhat) = \sum_{lm} (E_{lm} \pm i B_{lm}) {}_{\pm2}Y_{lm}(\vnhat),
\end{equation}
and orthogonality of the spin-weight spherical harmonics we can integrate Eq.~\eqref{QUfullsky} against ${}_{\pm2}Y_{lm}^*$ to extract the E and B modes. Expressing the integral of three spherical harmonics in terms of Wigner-3$j$ symbols
this gives the total B-mode coefficients (including the standard lensing effect~\cite{Okamoto:2003zw}) as
\begin{multline}
i\tilde{B}^{(2)}_{\ell m} = (-1)^{m+1} \sum_{LM,L'M'} \sqrt{\frac{(2L+1)(2L'+1)(2\ell+1)}{4\pi}}
\threej{L}{L'}{\ell}{M}{M'}{-m}
\threej{L}{L'}{\ell}{0}{-2}{2}\beta_{\ell LL'}\\
\times \biggl\{\sqrt{(L'+2)!/(L'-2)!}\psi_{d,L'M'} \psi_{\zeta,LM}
+ \psi_{d,LM} E_{L'M'}
-\frac{1}{2}\left[L(L+1)+L'(L'+1)-\ell(\ell+1)\right] \phi_{LM} E_{L'M'}
\biggr\},
\end{multline}
where $\beta_{\ell LL'} \equiv [1-(-1)^{\ell +L+L'}]/2$. The additional contribution to the full-sky lensing B-mode power spectrum from emission-angle and time-delay effects at leading order is then
\begin{multline}
\Delta\tilde C_{\ell}^{BB} =
\sum_{LL'} \frac{(2L+1)(2L'+1)}{4\pi}\beta_{\ell LL'}\\ \times
\biggl\{
\left[(L'+2)!/(L'-2)!C^{\psi_\zeta}_L C^{\psi_d}_{L'} + C^{EE}_{L'} \left(C^{\psi_d}_L
- C^{\psi_d \phi}_L \left[L(L+1)+L'(L'+1)-\ell(\ell+1)\right] \right)
\right]
\threej{L}{L'}{\ell}{0}{-2}{2}^2
\\
-C^{E\psi_{\zeta}}_{L}\left[2 C^{\psi_d}_{L'}
-C^{\phi\psi_d}_{L'}\left[L(L+1)+L'(L'+1)-\ell(\ell+1)\right]
\right]
\sqrt{(L'+2)!/(L'-2)!}\threej{L}{L'}{\ell}{0}{-2}{2}\threej{L'}{L}{\ell}{0}{-2}{2}
\biggr\}.
\label{FullSkyBB}
\end{multline}
On large scales ($\ell \ll L$ and $L'$) the terms on the middle line dominate and we can replace the Wigner 3$j$ symbol with its asymptotic form
\begin{equation}
\beta_{\ell LL'}\threej{L}{L'}{\ell}{0}{-2}{2}^2 \approx \beta_{\ell LL'}\frac{\delta^K_{L\pm (\ell-1),L'}}{8L},
\end{equation}
where $\delta^K_{ij}$ is the Kroneker delta. This implies a white-noise power spectrum with amplitude
\begin{align}
\Delta\tilde C_{\ell}^{BB} &\approx \sum_{L}\frac{L^3}{4\pi} \left(L^2 C^{\psi_\zeta}_L C^{\psi_d}_L  - 2C^{EE}_L C^{\phi \psi_d}_L\right)\\
&\approx \int \frac{L^3\ud L}{4\pi} \left(L^2 C^{\psi_\zeta}_L C^{\psi_d}_L  - 2C^{EE}_L C^{\phi \psi_d}_L\right),
\end{align}
which matches the flat-sky result given in Eq.~\eqref{BBtotalwhite}. Numerical results using Eq.~\eqref{FullSkyBB} also agree well with those presented in Sec.~\ref{sec:delay}.

%

\end{widetext}


\bibliography{PostBorn,antony,cosmomc}

\end{document}